\documentclass[prb,notitlepage,twocolumn]{revtex4-2}

\usepackage{graphicx}
\usepackage{amsmath}
\usepackage{hyperref}

\usepackage{yhmath}
\usepackage{orcidlink}
\usepackage{mathrsfs}

\begin{document}

\title{Dual Shapiro steps and fundamental transconductance
in the dc-driven Bloch transistor}
\author{A.~B.~Zorin\,\orcidlink{0000-0002-7706-1501}}
\affiliation{Independent Researcher}

\date{August 24, 2026}

\begin{abstract}
We propose a superconducting circuit based on the Bloch
transistor, a quantum device consisting of two
small-capacitance Josephson junctions, connected in series
and having a small superconducting island in between.
This device is driven by two dc electrical sources controlling
Josephson oscillations of frequency $f_J = 2e\overline{V_J}/h$,
related to the average transistor
voltage $\overline{V_J}$, and Bloch
oscillations of frequency $f_B = \overline{I_B}/2e$,
related to the average current $\overline{I_B}$ injected
into the transistor island.
We show that due to the Bloch transistor properties, these two
types of oscillations are coupled and can mutually phase-lock,
i.e., $f_J = f_B$.
This leads to the formation of a current step on the
current-voltage curve
at $\overline{I_B} = 2ef_J$, which is similar
to the dual Shapiro step,
appearing under microwave  irradiation of frequency $f$
on the current-voltage curve of a small
Josephson junction at current $I=2ef$.
Moreover, the Bloch transistor transconductance
$\overline{I_B}/\overline{V_J}$ takes the fundamental value
of $1/R_Q$, where $R_Q = h/4e^2$ is the resistance quantum.
The obtained results pave the way to an alternative quantum
standard of resistance, based on the superconducting circuit
and operating without applying strong magnetic field.
\end{abstract}
\maketitle

%%%%%%%%%%%%%%%%%%%%%%%%%%%%%%%%%%%%%%%%%%%%%%%%%%%
\section{Introduction}%%%%%%%%%%%%%%%%%%%%%%%%%%%%%
%%%%%%%%%%%%%%%%%%%%%%%%%%%%%%%%%%%%%%%%%%%%%%%%%%%

Significant progress in manufacturing small superconducting tunnel
junctions made it possible for the rapid growth of solid-state quantum
technologies including the development of Josephson qubits \cite{Pashkin2009,Koch2007,Clarke2008,Manucharyan2009}. These circuits
exploit the multiband structure of the energy spectrum enabling
quantum engineering of two-level systems \cite{MakhlinRMP2001}
with high fidelity \cite{Pechenezhskiy2020}.
The development of the so-called Bloch circuits
exploiting the $2e$-periodicity of the ground state was not,
however, so rapid.
As was predicted in Refs.\,\cite{DA-AZ-KKL-JETP1985,KKL-AZ-JLTP85},
these circuits, being fed by constant
current $I$, generate Bloch oscillations with frequency $f_B=I/2e$.
This phenomenon is associated with tunneling of single Cooper
pairs and it is dual to ac Josephson effect \cite{JOSEPHSON1962}.
Similar to Shapiro steps appearing on the I-V curves of
microwave-irradiated Josephson junctions, the ac-driven Bloch circuits
demonstrate on their I-V curves the so-called dual Shapiro steps
at $I=2nef$, where $f$ is the frequency of the external ac signal and
$n$ is an integer.
Slowdown of the experiments with Bloch oscillations can be
explained by serious problems in the implementation of compact
high-ohmic impedance in biasing circuits \cite{KuzminHaviland1991,KuzminHaviland1992,KuzminPashkinClaeson1994,
Pashkin1996,WatanabeHaviland2001}, i.e.
$|Z|\gg R_Q=h/4e^2 \approx 6.453\,\textrm{k}\Omega$,
where $R_Q$ is the resistance quantum, related to the von Klitzing
constant $R_K=h/e^2=4R_Q$.
Such high impedance is necessary for efficient suppression
of quantum fluctuations of charge in the frequency range up
to tens of GHz \cite{DA-AZ-KKL-JETP1985,KKL-AZ-JLTP85}.

Significant improvement of the Bloch circuits has been made using
high-quality ultra-high-ohmic microstripline resistors \cite{Lotkhov2013}
and the so-called superinductors. The latter elements are based on large
kinetic inductance of
either arrays of Josephson junctions \cite{Manucharyan2009} or
superconducting granular microstrips \cite{Grunhaupt2019}.
A large superinductor in combination with serial resistance
can substantially mitigate the problem of high-impedance implementation.
Recently, several demonstrations of dual Shapiro steps using the
superinductance concept
have been reported by the PTB-Braunschweig \cite{KaapPRL2024,KaapNatComm2024},
CNRS-Grenoble \cite{Crescini2023}, and RHUL \cite{ShaikhaidarovNatCom2024}
groups. Similar dual Shapiro steps were also demonstrated in the
experiment with microwave-driven quantum phase slip (QPS) circuits \cite{Shaikhaidarov2022,Antonov2026}.
Although the shape of the observed steps was not perfect (presumably
because of considerable noise, associated with high effective
electron temperature and/or interferences caused by microwave
irradiation) the positions of the step centers were
unambiguously related to the expected values of $I = 2nef$.
Motivated by the results of these remarkable experiments, we propose
an alternative Bloch-Josephson circuit enabling efficient phase lock
of Bloch oscillations and, hence, observation of dual Shapiro steps
\emph{without} applying an external microwave signal.

Our concept is based on the quantum properties of the
Bloch transistor (BT), a three-terminal device made of
two small serial Josephson junctions and a small superconducting
island in between having a control gate electrode. The BT
ground state is described by phase $\phi$ across the device and
quasicharge $q$ associated with the island
\cite{KKL1986preprint,KKL-AZ-JJAP1987RQ,AZ-PRL1996}.
Two electrical dc sources govern the dynamics of these variables;
in particular, they control the frequencies of Josephson ($f_J$)
and Bloch ($f_B$) oscillations, respectively.
When these frequencies are close to each other,
phase locking occurs, $f_J=f_B\equiv f$, and a step, similar to
the dual Shapiro step under microwave irradiation, appears
on the I-V curve of a dc-driven BT at $I = 2ef_J$.
Moreover, because $f_J$ and $f_B$ are related to voltage and current
via the fundamental constants, $\Phi_0 = h/2e$ and $2e$, respectively,
their phase locking results in quantization of the transconductance
in units of $R_Q^{-1}$.
Quantization of transconductance has been earlier found
in a two-dimensional electron gas (2DEG), realized in a silicon
field-effect transistor in a strong magnetic field \cite{KvKlitzingPRL1980}
and that discovery had made great impact on quantum
metrology  \cite{KvKlitzingRMP1986}.
Transconductance quantization was later predicted in multi-terminal
Josephson circuits \cite{Riwar2016, Eriksson2017}, but
to the best of our knowledge was not realized in experiment yet.

The idea of phase locking of the Bloch and Josephson oscillations
was first proposed by Likharev \cite{KKL1986preprint}.
It has been somewhat elaborated in Ref.\,\cite{KKL-AZ-JJAP1987RQ},
although without deep analysis and details of possible experiments.
To the best of our knowledge, this idea
was not experimentally realized, which was seemingly related
to insufficiencies in fabrication
technology and imperfections of low-noise cryogenic experiment
at that time.

Finally we note that an alternative approach to synchronization
of the Bloch and Josephson oscillators has been proposed for a QPS
circuit in Ref.\,\cite{HriscuNazarovPRL2013RQ}.
The authors suggested a nanowire-based QPS element coupled
to a \emph{stand-alone} Josephson oscillator by means of a
low-Q resonator.
Recently, an attempt to integrating \emph{individual} Josephson
and Bloch oscillators on one chip was reported
in Ref.\,\cite{ShaikhaidarovAPL2024}.
In contrast to these circuits, the proposed circuit has neither
a resonator (obviously shrinking a phase-locking range),
no a stand-alone Josephson oscillator as such.

%%%%%%%%%%%%%%%%%%%%%%%%%%%%%%%%%%%%%%%%%%%%%%%
\section{Electrical circuit}%%%%%%%%%%%%%%%%%%%
%%%%%%%%%%%%%%%%%%%%%%%%%%%%%%%%%%%%%%%%%%%%%%%

The schematic of our circuit is shown in Fig.\,1.
It consists of three parts: the Bloch circuit
driven by constant voltage source $V_0$, the Josephson
circuit, driven by constant current source $I_0$,
and the BT enabling coupling of these circuits.
Due to sufficiently low temperature and presumably
slow dynamics, BT remains in the ground state.
The parameters of the circuits are chosen such,
that the two basic variables of BT,
quasicharge $q$ and overall phase
$\phi = \varphi_1+\varphi_1$ \cite{KKL1986preprint,KKL-AZ-JJAP1987RQ},
where $\varphi_{1,2}$ are individual phase drops on the junctions,
are essentially decoupled from the environment and their
behavior is classical.

%%%%%%%%%%%%%%%%%%%%%%%%%%%%%%%%%%%%%%%%%%%%%%%%%%%%
\begin{figure}
\begin{center}
\includegraphics[width=3.4in]{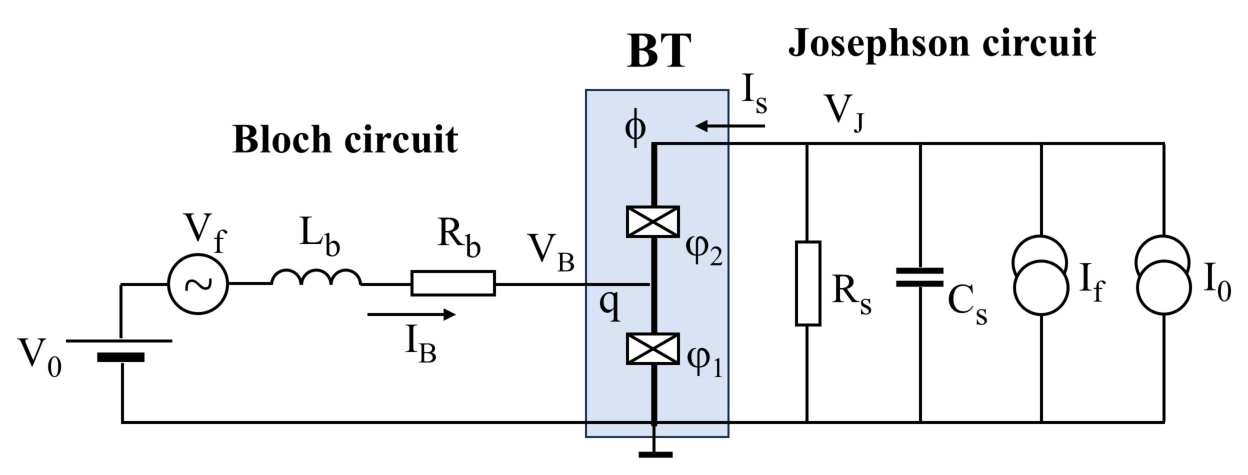}
\caption{
Schematic of the proposed circuit including Bloch transistor
(BT), outlined by a grey rectangle, and the two
complementary circuits, i.e., the Bloch and the Josephson
circuits.
The series Bloch circuit includes high-ohmic resistance $R_b$,
inductance $L_b$, voltage noise source $V_f$, and constant
voltage source $V_0$.
Current $I_B = dq/dt$ is injected into the BT island; the island
voltage $V_B$ depends on both quasicharge $q$ and total
phase $\phi$.
The parallel Josephson circuit comprises low-ohmic
resistance $R_s$, rather large capacitance $C_s$, current
noise source $I_f$, and constant current source $I_{0}$.
The BT voltage $V_J = d\phi/dt$, while supercurrent $I_s$
depends on both phase $\phi$ and quasicharge $q$.
}
\label{Fig:schematic1}
\end{center}
\end{figure}
%%%%%%%%%%%%%%%%%%%%%%%%%%%%%%%%%%%%%%%%%%%%%%%%%%%%

Resistance $R_b$, inserted in the Bloch circuit,
is assumed to be sufficiently large
\cite{DA-AZ-KKL-JETP1985,KKL-AZ-JLTP85}, so the corresponding
dimensionless parameter is small,
\begin{equation}
\alpha_B = R_Q/R_b \ll 1.  \label{alfa-B}
\end{equation}
Resistance $R_b$ is compact, so its stray capacitance is small
\cite{Lotkhov2013,KaapNatComm2024}.
Optional series (super)inductance $L_b$, like in recent experiments
of the RHUL group \cite{ShaikhaidarovNatCom2024,Antonov2026}, can
notably suppress quantum fluctuations of charge on the BT island
at high frequencies.

The Josephson part of the circuit includes the
low-ohmic parallel resistance $R_s$, so the corresponding
small dimensionless parameter is
\begin{equation}
\alpha_J = R_s/R_Q \ll 1.  \label{alfa-J}
\end{equation}
The parallel capacitance $C_s$ includes stray capacitances
of resistor $R_s$ and of the wires connecting
BT to source $I_0$. The shunting resistance $R_s$
is responsible for unwanted fluctuations of phase $\phi$ (see, for
example, Chapters 3 and 4 in Ref.\,\cite{KKLikharev-book}), so
the value of $R_s$ should be as low as possible.
The physical dimensions of this resistance are practically
limited by the size of the chip packaging, while the value of
unavoidable capacitance $C_s$ is not critical. So, we will
assume that $\alpha_J$ is very small,
\begin{equation}
\alpha_J\ll \alpha_B \ll 1.  \label{alfaJ-alfaB}
\end{equation}
For typical experimental parameters of BT, having
effective critical current $I_c^*$ on the order of 10-100~nA \cite{KaapNatComm2024,ShaikhaidarovNatCom2024},
and, say, resistance $R_s \sim 1\,\Omega$,
even huge capacitance $C_s\sim 1\,$nF gives small value
of the effective McCumber parameter
\cite{McCumber,Stewart},
$\beta_c=(2\pi/\Phi_0)I_c^* R_s^2C_s\lesssim 1$.
Thus, the effective Josephson junction (in fact, the BT) is
overdamped and the effect of capacitance $C_s$
on its dynamics is indeed small \cite{KKLikharev-book}.

Applying Kirchhoff's circuit laws, we obtain a set of two
coupled equations, for the Bloch and Josephson
circuits, respectively,
\begin{eqnarray}
L_b \frac{d^2q}{dt^2} + R_b \frac{dq}{dt} + V_B(q,\phi) = V_0 + V_f  ,~ \label{Eq-motion1}\\
\frac{\Phi_0}{2\pi}  C_s \frac{d^2\phi}{dt^2} + \frac{\Phi_0}{2\pi} R_s^{-1}
\frac{d\phi}{dt} + I_s(q,\phi) = I_0 + I_f.~\label{Eq-motion2}
\end{eqnarray}
Here we used the fact that gate current $I_B$ and the Josephson
voltage $V_J$ (see Fig.\,1)
are related to time derivatives of corresponding variables
\cite{DA-AZ-KKL-JETP1985,JOSEPHSON1962},
\begin{equation}
I_B = dq/dt~~\textrm{and}~~ V_J= (\Phi_0/2\pi) d\phi/dt. \label{IB-VJ}
\end{equation}
Gate voltage $V_B$ and Josephson supercurrent $I_s$ depend
on both variables, $q$ and $\phi$, and, hence, are responsible for
their coupling.
The shape of these double-periodic functions is determined by the BT
parameters. The corresponding formulas will be derived in Sec. III B.

The last terms on the right hand
sides of Eqs.(\ref{Eq-motion1}) and (\ref{Eq-motion2}),
variables $V_f$ and $I_f$, describe fluctuations in
the Bloch and Josephson circuits, respectively.
In the general case, their spectral densities are related to the
real parts of the corresponding impedance and admittance \cite{DA-AZ-KKL-JETP1985,KKL-AZ-JLTP85},
\begin{eqnarray}
S_V(\omega)=\frac{R_b}{\pi} \frac{\hbar \omega}{2}
\coth \frac{\hbar \omega}{2k_BT}, \label{Sv-1} \\
S_I(\omega)=\frac{1}{\pi R_s} \frac{\hbar \omega}{2}
\coth \frac{\hbar \omega}{2k_BT}.
\label{SvSi}
\end{eqnarray}
In order to keep our analysis simple, the telegraph noise,
associated with random appearance and disappearance of
quasiparticles on the island  \cite{AumentadoPRL2004},
i.e., poisoning of the island, is not included in our model.
So, we assume that in the experiment, the chip includes,
for example, appropriate quasiparticle
traps \cite{Nguyen_2013} or has metallization of its back side
\cite{Iaia2022}, so the island is free of non-equilibrium
quasiparticles.

%%%%%%%%%%%%%%%%%%%%%%%%%%%%%%%%%%%%%%%%%%%%%%%
\section{Bloch Transistor}%%%%%%%%%%%%%%%%%%%
%%%%%%%%%%%%%%%%%%%%%%%%%%%%%%%%%%%%%%%%%%%%%%%

The analysis of this remarkable device has been earlier
reported, for example, in the papers about the BT-based
electrometers \cite{AZ-PRL1996,ZorinPRL2001} and the
charge-phase qubits \cite{Vion2002,AZ-PhysC2002,AZ-JETP2004}.
Briefly, the Josephson junction free energies
are \cite{Barone-Paterno-book}
\begin{equation}
\mathscr{E}_{1,2}(\varphi_{1,2})=E_{J1,J2}(1-\cos\varphi_{1,2}).
\end{equation}
We admit that the Josephson energies, $E_{J1,J2}=(\Phi_0/2\pi)I_{c1,c2}$,
and, hence, critical currents $I_{c1,c2}$, can be unequal.
The sum of the two junction energies can be presented
as \cite{AZ-PRL1996}
 \begin{equation}
\mathscr{E}_1+\mathscr{E}_2 =
- E_J(\phi)\cos[\varphi + \gamma(\phi)]+(E_{J1}+E_{J2}),  \label{E-sum}
\end{equation}
where the last term is an unimportant constant, while
\begin{eqnarray}
 E_J(\phi) = (E_{J1}^2+E_{J2}^2+2E_{J1}E_{J2}\cos\phi)^{1/2},
 \label{E_J-vs-phi} \\
\phi = \varphi_1+\varphi_2, \label{phi}  \\
\varphi = (\varphi_1-\varphi_2)/2, \label{varphi} \\
 \gamma(\phi) =\arctan\left( a \tan\phi/2\right),  \label{gamma-phi}
\end{eqnarray}
and the asymmetry parameter
\begin{equation}
a =(E_{J1}-E_{J2})/(E_{J1}+E_{J2}).  \label{asymm}
\end{equation}
Then the ratio of the critical currents of individual
junctions is
\begin{equation}
r = I_{c2}/I_{c1}=E_{J2}/E_{J1}=(1-a)/(1+a).  \label{asymm r}
\end{equation}
We assume that $r \leq 1$, or, equivalently, $a\geq0$,
for definiteness.

%%%%%%%%%%%%%%%%%%%%%%%%%%%%%%%%%%%%%%%%%%%%%%%%%%%%%%%%%%%
\subsection{Hamiltonian }%%%
%%%%%%%%%%%%%%%%%%%%%%%%%%%%%%%%%%%%%%%%%%%%%%%%%%%%%%%%%%%

In the case of a low parallel impedance inserted into
the Josephson circuit, for example, by means of a large
capacitance $C_s$ ($e^2/2C_s \ll E_{J1,J2}$), associated
with an additional large Josephson junction parallel to
BT \cite{Vion2002}, or a small
superconducting inductance shunting BT \cite{AZ-PhysC2002,ZorinPRL2001},
or a small shunting resistance \cite{AZ-PRL1996}, as in our case,
phase $\phi$ is classical.
(The BT characteristics in the opposite case of a high external
impedance were reported in
Refs.\,\cite{Pashkin1996,AZ_Pashkin1998,AZ-1999-BTelectrometers,houzet2026blochdiode}.)
Then capacitances of individual Josephson junctions $C_{1,2}$
sum up with stray capacitance $C_{\textrm{stray}}$
and give total capacitance of the island
$C_{\Sigma} = C_1+C_2+ C_{\textrm{stray}}$.
As a result, the charging energy of BT is equal to
\begin{equation}
 E_c=e^2/2C_\Sigma. \label{Echarge}
\end{equation}
The  semidifference phase $\varphi$  Eq.\,(\ref{varphi})
is associated with the island and in the case of nonzero
charging energy $E_c$, it behaves quantum mechanically.
Variable $\varphi$ is conjugate to the island charge variable,
$Q = -2ie \partial/\partial\varphi$,
and obeys the commutation relation, $[\varphi,Q]=2ie$ \cite{DA-AZ-KKL-JETP1985}.

The BT Hamiltonian for fixed $\phi$ and zero gate current
(i.e., the Cooper pair box configuration
of BT \cite{Bouchiat1998,Pashkin2009})
takes the form \cite{KKL1986preprint,AZ-PRL1996}
\begin{equation}
H = \frac{Q^2}{2C_{\Sigma}}
- E_J(\phi)\cos[\varphi + \gamma(\phi)],  \label{Ham}
\end{equation}
yielding Schr\"{o}dinger equation (in fact, the Mathieu equation):
\begin{equation}
\left[ 4E_c \frac{\partial^2}{\partial\chi^2}
+ E_J(\phi)\cos\chi +E \right]\psi=0. \label{SchrEqu}
\end{equation}
Taking advantage of the relation $\partial/\partial \varphi
= \partial/\partial \chi$,
we introduce here a new variable,
\begin{equation}
\chi=\varphi+\gamma(\phi). \label{q-chi}
\end{equation}
Then the phases on individual junctions are:
\begin{equation}
\varphi_1=\chi-\gamma + \phi/2~~\textrm{and}~~
\varphi_2 =-\chi+\gamma + \phi/2.    \label{phi1-via-gamma}
\end{equation}

\begin{figure}
\begin{center}
\includegraphics[width=3.2in]{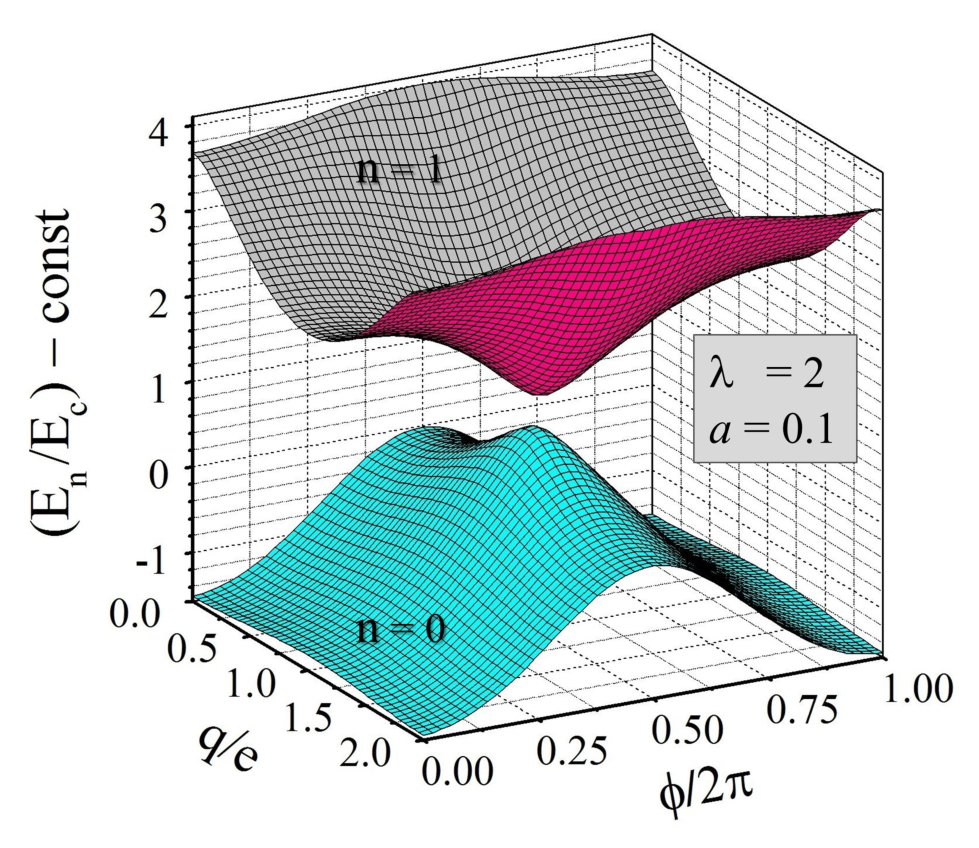}
\caption{
Elementary cell ($0 \leq q \leq 2e~\textrm{and}~0\leq
\phi \leq 2\pi$)
of the double-periodic surfaces of the
two lowest energy bands, i.e., the ground state band ($n = 0$)
and the first band ($n = 1$). The specific shape of these Bloch
bands has been used for engineering the charge-phase qubits
with inherently weak decoherence due to operating in the
so-called sweet points (in the extrema and saddle points)
\cite{Vion2002,AZ-JETP2004}.}
\label{Fig:3D2bands}
\end{center}
\end{figure}

Eigenvalue equation (\ref{SchrEqu}) yields the
Bloch bands $E_n(q,\phi)$ with band
index $n=0,1,2...$. These energies depend on quasicharge $q$
(good variable) \cite{KKL-AZ-JLTP85},
while $\phi$ is a real parameter \cite{KKL1986preprint}.
For any $n$, even function $E_n(q,\phi)$,
\begin{equation}
 E_n(-q,\phi) =  E_n(q,\phi)~~\textrm{and}
 ~~E_n(q,-\phi)= E_n(q,\phi), \label{En-symmetry}
\end{equation}
obeys the double periodicity condition,
\begin{equation}
 E_n(q,\phi) =  E_n(q+2e,\phi)
 = E_n(q,\phi+2\pi). \label{En-periodicity}
\end{equation}
The shape of the energy bands crucially depends on
the characteristic energy ratio,
\begin{equation}
\lambda = (E_{J1}+E_{J2})/2E_c, \label{EJEc-ratio}
\end{equation}
and the asymmetry parameter $a$ Eq.(\ref{asymm}).
As an illustration, a 3D plot of the energy surfaces for the
zeroth and the first Bloch bands is shown in Fig.\,\ref{Fig:3D2bands}.
%%%%%%%%%%%%%%%%%%%%%%%%%%%%%%%%%%%%%%%%%%%%%%%%%%%%
%%%%%%%%%%%%%%%%%%%%%%%%%%%%%%%%%%%%%%%%%%%%%%%%%%%%
Quantitatively, the shapes of the Bloch bands can be described
by infinite matrices $\textbf{A}^{(n)}$ with real elements
\begin{equation}
A_{km}^{(n)} = -\frac{1}{\pi e E_c} \int_{0}^{\pi}\int_{0}^{e}
E_n(q,\phi) \cos\frac{k \pi q}{e}\cos (m \phi) \,dq d\phi,~~ \label{Anm}
\end{equation}
entering the corresponding double Fourier series as
coefficients (see Appendix A).

The eigenstates given by equation (\ref{SchrEqu}) are the
Bloch waves,
\begin{equation}
\psi^{(n,\phi)}_q(\chi)
= w^{(n,\phi)}_q(\chi) e^{i\chi q/2e}, \label{BlochWaves}
\end{equation}
where $w^{(n,\phi)}_q(\chi)$ are the wave amplitudes, $q$ is quasicharge,
and $ q/2e =\kappa$ is the appropriate "crystal momentum" \cite{KKL-AZ-JLTP85}.

%%%%%%%%%%%%%%%%%%%%%%%%%%%%%%%%%%%%%%%%%%%%%%%%%%%%%%%%%%%
\subsection{Voltage $V_B$ and current $I_s$}%%
%%%%%%%%%%%%%%%%%%%%%%%%%%%%%%%%%%%%%%%%%%%%%%%%%%%%%%%%%%%

In our analysis, we focus on the ground state
$E_0(q,\phi)$, assuming that the bath temperature $T$ is
low, i.e.,
\begin{equation}
k_BT \ll E_1(q,\phi)-E_0(q,\phi),
\end{equation}
for all relevant $q$ and $\phi$.
Moreover, we assume that the time variations of
quasicharge $q$ and phase $\phi$ are sufficiently slow,
so the system remains in its eigenstate [given by
Hamiltonian (\ref{Ham})] at any time instant.
Naturally, we exclude unwanted transitions to the higher
bands due to this motion.
In the general case, phase operator $\chi$ has nonzero both
intraband and interband components \cite{LifshitzPitaevskii-book}.
However, for a fixed band index, $n=0$, one can neglect the
interband part of $\chi$ and consider that $[\chi, q]= 2ie$
\cite{DA-AZ-KKL-JETP1985,KKL-AZ-JLTP85}.

The electrical potential of the transistor island
with respect to ground (see Fig.~\ref{Fig:schematic1}),
\begin{equation}
V_B = (\Phi_0/2\pi) \langle d \varphi_1/d t\rangle,\label{V-phi1}
\end{equation}
can be found using Eq.(\ref{phi1-via-gamma}),
i.e.,
\begin{equation}
\frac{2\pi}{\Phi_0}V_B=\langle\frac{d\varphi_1}{d t}\rangle=
\langle\frac{\partial \chi}{\partial t}\rangle
- \frac{\partial\gamma}{\partial\phi}
\frac{\partial \phi}{\partial t} +
\frac{1}{2} \frac{\partial \phi}{\partial t}, \label{dphi1-dt}
\end{equation}
where $\langle...\rangle$ denotes ensemble averaging.
The first term on the right hand side of Eq.(\ref{dphi1-dt})
is determined by the shape of the ground state
energy \cite{DA-AZ-KKL-JETP1985,KKL-AZ-JLTP85},
\begin{equation}
\frac{\Phi_0}{2\pi}\langle \frac{\partial \chi}{\partial t}\rangle
= \frac{\partial E_0(q,\phi)}{\partial q}
\equiv V_B^{(0)}(q,\phi).\label{voltage-dEdq}
\end{equation}
This formula yields voltage $V_B^{(0)} = V_B$ for
fixed $\phi$, i.e., when $\partial\phi/\partial t =0$ and
both the second and the third terms in Eq.(\ref{dphi1-dt}) are zero.
Note that the island voltage derived (equivalently!) using phase $\varphi_2$,
i.e., $V_B = -(\Phi_0/2\pi)\langle \partial \varphi_2/\partial t \rangle
+ V_J$, where voltage drop $V_J=(\Phi_0/2\pi)\partial\phi/\partial t$
(see Fig.~1), is given by the same formula Eq.(\ref{dphi1-dt}).

Using expression Eq.(\ref{gamma-phi}) for $\gamma$, both
the second and the third terms in Eq.(\ref{dphi1-dt}) can
be rewritten in the form:
\begin{equation}
- \frac{\partial\gamma}{\partial\phi}
\frac{\partial \phi}{\partial t} +
\frac{1}{2} \frac{\partial \phi}{\partial t}= -g(\phi)
\frac{\partial \phi}{\partial t},
\end{equation}
where the $2\pi$-periodic function $g(\phi)$,
\begin{equation}
g(\phi) = \frac{0.5a}{\cos^2(\phi/2)+a^2 \sin^2(\phi/2)}-0.5, \label{g-phi}
\end{equation}
has zero average value, i.e. $\int_0^{2\pi} g(\phi)d\phi=0$.
Summarizing, we obtain formula
\begin{equation}
V_B(q,\phi) = V_B^{(0)}(q,\phi)
-\frac{\Phi_0}{2\pi}g(\phi)\frac{\partial \phi}{\partial t}, \label{V-B-V(0)}
\end{equation}
where the second, dynamical term ($\propto d\phi/dt \neq 0$) is
vanishingly small, for example, in the trivial case of a highly
asymmetric BT, i.e., $r \rightarrow 0$ (or $a\rightarrow 1$).

Similar to ground state energy $E_0(q,\phi)$, voltage
$V_B^{(0)}(q,\phi)$ Eq.(\ref{voltage-dEdq}) is also
double periodic,
\begin{equation}
 V_B^{(0)}(q,\phi) =  V_B^{(0)}(q+2e,\phi)
 = V_B^{(0)}(q,\phi+2\pi). \label{V-periodicity}
\end{equation}
This function is odd (even) with respect to quasicharge $q$
(phase $\phi$):
\begin{equation}
 V_B^{(0)}(-q,\phi)= -V_B^{(0)}(q,\phi)~~\textrm{and}~~
 V_B^{(0)}(q,-\phi)=V_B^{(0)}(q,\phi). \label{V-symmetry}
\end{equation}
Voltage $V_B^{(0)}$ is controlled by quasicharge $q$, determined by
the injected current $I_B$ \cite{DA-AZ-KKL-JETP1985},
\begin{equation}
q = q_0 + \int_0^t I_B(t') dt',  \label{q-vs-IB}
\end{equation}
and apparently by phase $\phi$.

Josephson supercurrent $I_s$ flowing through BT also
depends on these two variables. It can be expressed as an
ensemble average of the time derivative of the corresponding
charge operator, $Q_s = -2ie \partial/\partial \phi$.
This operator is conjugate to the phase operator $\phi$.
Thus the supercurrent is
\begin{equation}
I_s = \langle d Q_s/dt \rangle.  \label{Is-via-Q}
\end{equation}
Using the Heisenberg equation of motion for operators, we obtain
\begin{equation}
I_s = -2ie \langle \frac{d }{dt} \frac{\partial}{\partial \phi} \rangle
 = \frac{2\pi}{\Phi_0} \langle [H,\frac{\partial}{\partial \phi}]\rangle
 = \frac{2\pi}{\Phi_0}\frac{\partial E_0(q,\phi)}{\partial \phi}.   \label{Is-via-Q2-2}
\end{equation}
Being a function of two classical variables, $q$
and $\phi$, supercurrent $I_s(q,\phi)$ is an odd (even) function of
phase $\phi$ (quasicharge  $q$), possessing double periodicity, i.e.,
\cite{AZ-PRL1996},
\begin{eqnarray}
I_s(q,-\phi) =  -I_s(q,\phi)~~\textrm{and}~~I_s(-q,\phi)
= I_s(q,\phi),~~\label{Is-symmetry} \\
I_s(q,\phi)= I_s(q,\phi+ 2\pi)=I_s(q +2e,\phi).~~\label{Is-periodicity}
\end{eqnarray}
The current-phase relation, $I_s$ versus $\phi$,
is non-sinusoidal for any $q$ \cite{AZ-PRL1996}
(see Eq.(\ref{i1i2}) in Appendix A) and the critical current value,
\begin{equation}
I_c(q) = \max_{\phi \in [0,\pi]} I_s(q,\phi), \label{Ic-BT}
\end{equation}
is always smaller than the value in the double Josephson junction in the
absence of the Coulomb blockade effect on the island ($E_c \rightarrow 0$) \cite{KKL1986preprint,Matveev1993,AZ-PRL1996}. The periodic dependence
of critical current $I_c$ on quasicharge $q$ was reported earlier, for
example, in Refs.\cite{Joyez1994, EilesMartinis1994}.
In these experiments, $q$ was controlled by a voltage applied to the
capacitively coupled gate.

The expressions for functions $E_0(q,\phi)$, $V_B^{(0)}(q,\phi)$,
and $I_s(q,\phi)$ in the form of the double Fourier series are given
in Appendix A.

%%%%%%%%%%%%%%%%%%%%%%%%%%%%%%%%%%%%%%%%%%%%%%%
\section{Equations of Motion}%%%%%%%%%%%%%%
%%%%%%%%%%%%%%%%%%%%%%%%%%%%%%%%%%%%%%%%%%%%%%%

We rewrite Eqs. (\ref{Eq-motion1}) and (\ref{Eq-motion2}) in
the dimensionless form using the normalizing units
\begin{equation}
V_u=(\pi/2) e/C_\Sigma ~~~\textrm{and}~~~I_u=V_u/R_Q \label{V_u-norm}
\end{equation}
for the voltage and the current, respectively.
The dimensionless quasicharge
is defined as
\begin{equation}
\theta=\pi q/e,   \label{theta-q}
\end{equation}
and the time derivatives are now
denoted by a dot over the variables.
Due to the periodic terms $v_B$ and $i_c$, the resulting
oscillator equations
for the Bloch and Josephson circuits (cf. with
Eqs.\,(\ref{Eq-motion1}) and (\ref{Eq-motion2})),
\begin{eqnarray}
\omega_{Bp}^{-2} \ddot{\theta} +\omega_{Bc}^{-1}\dot{\theta}
+ v_B(\theta,\phi)= v_0 + v_f  ,~~ \label{Eq-motion1Norm}\\
\omega_{Jp}^{-2} \ddot{\phi}+\omega_{Jc}^{-1} \dot{\phi}
+ i_s(\theta,\phi) = i_0
+ i_f,~~\label{Eq-motion2Norm}
\end{eqnarray}
mimic the RSJ model equations \cite{McCumber,Stewart}.
Here the dimensionless BT voltage $v_B$ includes
two terms, corresponding to two terms in Eq.(\ref{V-B-V(0)}),
\begin{equation}
v_B(\theta,\phi)= v_B^{(0)}(\theta,\phi)
- \omega_{Bc}^{-1} \alpha_B g(\phi) \dot{\phi}. \label{vB-vB0}
\end{equation}
The corresponding "plasma" and "characteristic" frequencies in
the Bloch and Josephson equations are defined as
\begin{equation}
\omega_{Bp} = \frac{\pi}{\sqrt{2L_b C_\Sigma}},~~
\omega_{Bc}=
\alpha_B \left(\frac{2\pi}{\Phi_0}\right) V_u,\label{omega-B}
\end{equation}
and
\begin{equation}
\omega_{Jp}= \frac{\pi}{2R_Q \sqrt{C_\Sigma C_s}},~~
\omega_{Jc}
=\alpha_J \left(\frac{2\pi}{\Phi_0}\right) V_u,\label{omega-J}
\end{equation}
respectively.

%%%%%%%%%%%%%%%%%%%%%%%%%%%%%%%%%%%%%%%%%%%%%%%%%%%%
\begin{figure}
\begin{center}
\includegraphics[width=3.2in]{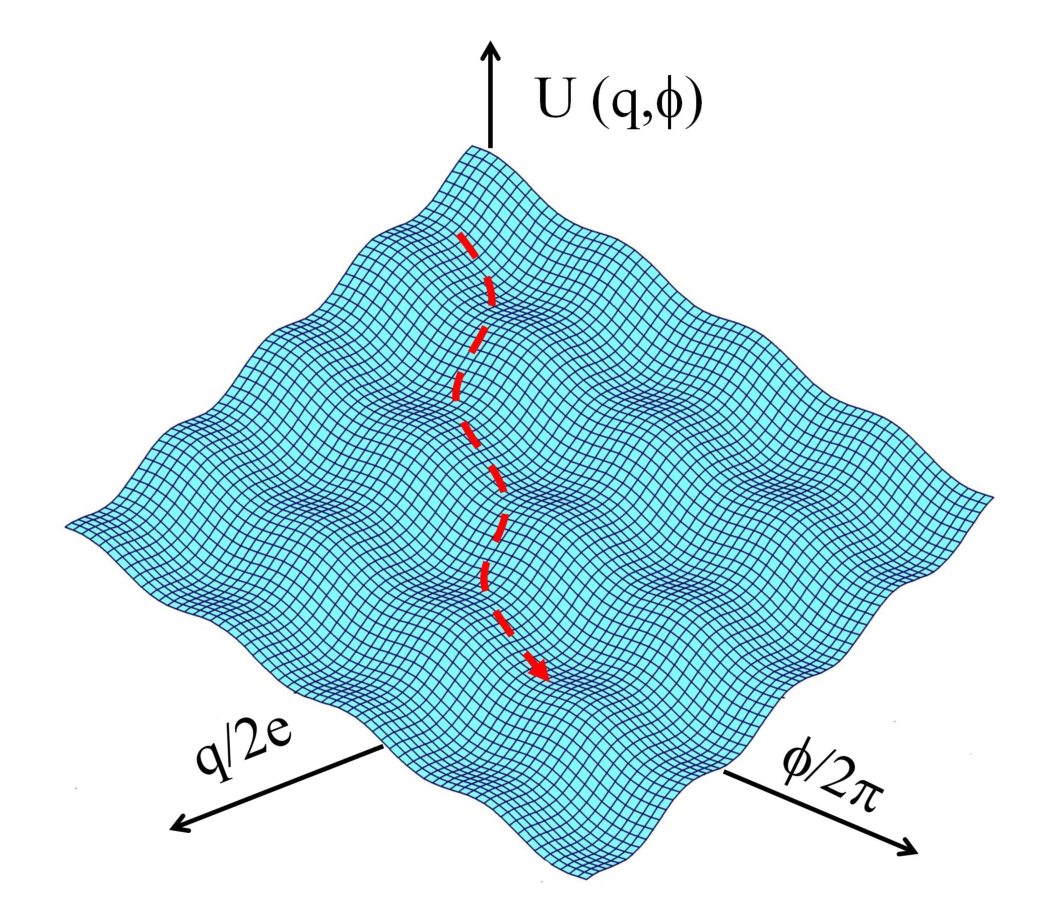}
\caption{Egg box potential Eq.(\ref{U-function}) with adjusted
individual slopes for variables
$q = \theta e/\pi$ and $\phi$, ensuring
motion downwards with constant average speed
along "diagonal" path (schematically
shown by the red dashed line). This path is determined by
the steepest downward slope in each point, while the motion is associated
with the circuit operation, when a working point is fixed on the step.
\label{Fig:egg-box}}
\end{center}
\end{figure}
%%%%%%%%%%%%%%%%%%%%%%%%%%%%%%%%%%%%%%%%%%%%%%%%%%%%

\subsection{Egg box potential}%%%

The double periodic functions $v_B^{(0)}(\theta,\phi)$
and $i_s(\theta,\phi)$
in combination with constant voltage $v_0$ and constant
current $i_0$, included in the right-hand sides of
Eqs.\,(\ref{Eq-motion1Norm}) and (\ref{Eq-motion2Norm}),
can be expressed via the derivatives of the dimensionless
potential energy $u(\theta,\phi,v_0,i_0)$. Specifically,
\begin{eqnarray}
v_B^{(0)}(\theta,\phi) - v_0
= \partial u/\partial \theta, ~\label{dU-dq} \\
i_s(\theta,\phi) - i_0 = \partial u/\partial \phi, ~\label{dU-dphi}
\end{eqnarray}
where
\begin{equation}
u= U(\theta,\phi,v_0,i_0)/E_c = \varepsilon(\theta,\phi)-\theta v_0 -\phi i_0.  \label{U-function}
\end{equation}
Here $\varepsilon(\theta,\phi) = E_0(q,\phi)/E_c$ is the normalized
ground state energy. Potential energy
$U(\theta,\phi)$ Eq.(\ref{U-function})
is a two-dimensional analog of the well-known washboard
potential playing a pivotal role in the physics of Josephson
junctions \cite{KKLikharev-book}. In our case, the cosine
Josephson potential is replaced by two-dimensional
surface $\varepsilon(\theta,\phi)$ with double-periodic extrema
and saddles. The shape of this surface mimics the shape of
an egg box. Voltage $v_0$ and current $i_0$ are associated with
the slopes for variables $q$ and $\phi$, respectively.

Figure~\ref{Fig:egg-box} shows the egg box potential with a
sufficiently large slope (both $v_0$ and $i_0$ are above
corresponding thresholds)
to keep running states for both variables.
%$\overline{d\theta/dt} > 0$ and $\overline{d\phi/dt} > 0$.
It is intuitively clear that for a certain relation between
$v_0$ and $i_0$ the synchronous motion of the two variables
is possible along the "diagonal" path, schematically shown
by the red dashed line.
Some tolerance of this running state to small variations
of $v_0$ (or  $i_0$) can be interpreted as a finite
phase-locking range (i.e., a finite size of the step).

%%%%%%%%%%%%%%%%%%%%%%%%%%%%%%%%%%%%%%%%%%%%%%%
\subsection{Operating Frequency}%%%%%%%%%%%%%%
%%%%%%%%%%%%%%%%%%%%%%%%%%%%%%%%%%%%%%%%%%%%%%%

Let us make a comment concerning the operating frequency
of the circuit, $\omega \sim \omega_B \sim \omega_J$,
that is controlled by $v_0$ and $i_0$ entering into
the right-hand sides of
Eqs.\,(\ref{Eq-motion1Norm}) and (\ref{Eq-motion2Norm}),
respectively.
Taking into account strict inequality (\ref{alfaJ-alfaB}), we
conclude that frequency $\omega_{Jc} \ll \omega_{Bc}$.
We will assume that operating frequency $\omega \gg \omega_{Jc}$
and this inequality allows us to simplify solving the system of
equations (\ref{Eq-motion1Norm}) and (\ref{Eq-motion2Norm}).
Specifically, the nonlinear term $i_s(\theta,\phi)$ in
Eq.\,(\ref{Eq-motion2Norm}) is much smaller than the second
(dominant) term,
$\omega_{Jc}^{-1} \dot{\phi} \sim \omega/\omega_{Jc} \gg 1$,
hence, $i_s$ can be considered as a small perturbation.
Thus, backaction of the Bloch circuit
on the Josephson circuit is negligible in the first order.

Note that in the physical phenomenon of phase locking of two
interacting generators, the regime when one device dominates
the other is quite possible \cite{Migulin-book}.
In our case, the slave Bloch oscillator is driven by the
steady Josephson oscillator and such choice of the circuit
parameters seems to be quite natural.
Using the typical island capacitance
$C_\Sigma \sim 1$\,fF and taking the resistance value
$R_s \sim 1\,\Omega$, we arrive at the Josephson
characteristic frequency of $\omega_{Jc}/2\pi \sim 20$\,MHz
[see Eqs.\,(\ref{alfa-J}), (\ref{V_u-norm}) and (\ref{omega-J})].
As will be shown below [see Eq.\,(\ref{gamma1})], the
differential resistance of the Bloch circuit at such
low frequencies is rather high and the effect of noise
is large.
Thus, the beautiful case of truly mutual interaction of
the Bloch and Josephson oscillators, occurring at low
operating frequency, $\omega \sim \omega_{Jc}$,
is seemingly of pure academic interest.

%%%%%%%%%%%%%%%%%%%%%%%%%%%%%%%%%%%%%%%%%%%%%%%%%%%%%%%%
\section{Phase locking}
%%%%%%%%%%%%%%%%%%%%%%%%%%%%%%%%%%%%%%%%%%%%%%%%%%%%%%%%

As long as the operating frequency is relatively high,
$\omega \gg \omega_{Jc} \sim (2\pi/\Phi_0)R_sI_c(q)$,
the Josephson circuit dynamics (\ref{Eq-motion2Norm}) is
described by the RSJ model in the high-frequency limit
\cite{KKLikharev-book}:
\begin{equation}
\phi = \omega t+ \widetilde{\phi} +\phi_0+
\phi_f,~~~~\omega=\omega_J=\frac{2\pi}{\Phi_0}\overline{V_J}
\approx \frac{2\pi}{\Phi_0}I_0 R_s, \label{phi-linear}
\end{equation}
where we can neglect the small oscillating term $\widetilde{\phi}$
and fix initial phase $\phi_0=0$. Noise variable $\phi_f$ obeys
the stochastic equation,
\begin{equation}
\omega_{Jp}^{-2}\dot{\phi}_f+\omega_{Jc}^{-1}\dot{\phi}_f
= i_f, \label{phi-f}
\end{equation}
with Johnson noise term $i_f$ on the right-hand side.
When $\omega_{Jp} \gg \omega_{Jc}$, i.e., in the case of
relatively small capacitance $C_s$ ($\beta_c \ll 1$),
Eq.\,(\ref{phi-f}) is reduced to equation
\begin{equation}
\dot{\phi}_f =  \omega_{Jc} i_f,~~\label{phi-Wiener}
\end{equation}
describing Brownian motion of phase $\phi_f$, having a variance
that increases linearly with time (see Eqs. (2.6.1a) and (2.6.8a)
in Ref. \cite{Axmanov-book}).
Substituting relation  Eq.\,(\ref{phi-linear})
into Eq.\,(\ref{Eq-motion1Norm})
yields the equation solely for dimensionless quasicharge $\theta$:
\begin{eqnarray}
\omega_{Bp}^{-2} \ddot{\theta}+\omega_{Bc}^{-1} \dot{\theta}
&+&v_B^{(0)}(\theta,\omega t + \phi_f) \nonumber \\
&=& \omega \omega_{Bc}^{-1} \alpha_B\, g(\omega t + \phi_f)+ v_0 +v_f.~~ \label{Eq-motion1-r}
\end{eqnarray}

\subsection{Autonomous I-V curve}%%%

Let us make several assumptions simplifying
solving Eq.(\ref{Eq-motion1-r}):

(i) Assuming that $\omega_{Bp} \gg \omega_{Bc}$
(the case of relatively low inductance $L_b$), we omit the
first term on the left-hand side of Eq.\,(\ref{Eq-motion1-r}).

(ii) In both sums in the Fourier series representation
of $v_B^{(0)}$ (\ref{v1v2}), we keep only
the first (dominant) terms, i.e.,
\begin{equation}
v_B^{(0)}(\theta,\phi)
=  u_c \sin\theta\, (1+\mu \cos\phi), ~~\label{v0-approx}
\end{equation}
where $u_c = 0.5 A_{10} = V_c/V_u$, $V_c = 0.25\pi A_{10}e/C_{\Sigma}$,
and $\mu =2 |A_{11}|/A_{10}$.

(iii) Function $g(\omega t)$  contains the
main tone $\omega$ and its harmonics $2\omega$, $3\omega$,...
[see the shape of function $g$ given by formula (\ref{g-phi})].
It is easy to show that the component oscillating with
frequency $\omega$ is \cite{Supp_Info_Zenodo2026}
\begin{equation}
[g(\phi)]_{\omega} = r \cos \phi, ~~\label{g-approx}
\end{equation}
where $r$ is defined by Eq.\,(\ref{asymm r}).
In fact, this term plays the role of small ac drive
in Eq.\,(\ref{Eq-motion1-r}).
As long as coefficient $\alpha_B \ll \omega/\omega_{Bc}\sim 1$ is
small, we can neglect that term.

\begin{figure}
\begin{center}
\includegraphics[width=3.2in]{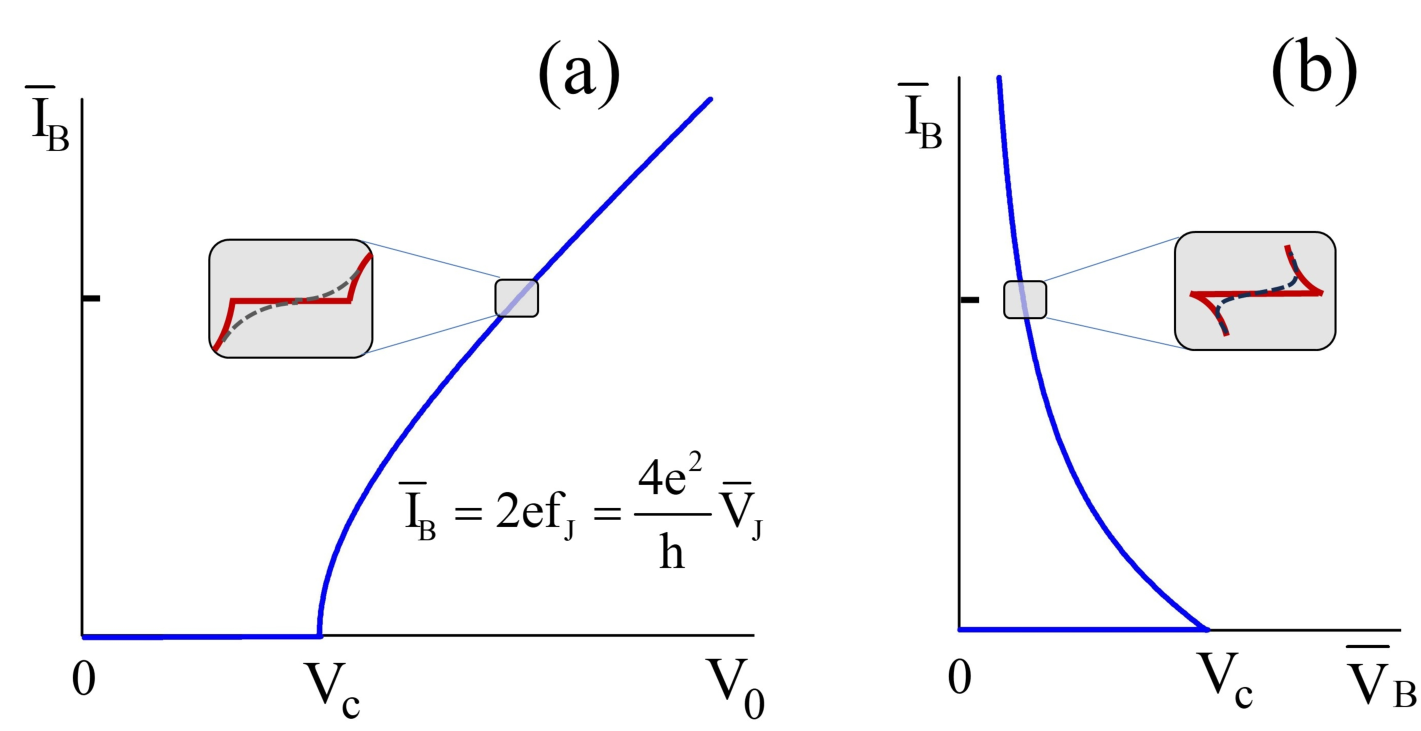}
\caption{Autonomous I-V curves of the Bloch circuit:
(a) average current $\overline{I_B}$ vs voltage $V_0$
and (b) the same curve replotted
as $\overline{I_B}$ against $\overline{V_B}$.
The insets show the dual Shapiro steps of constant current
in the "embryo stage" ($\mu \rightarrow 1$), appearing at
$\overline{I_B} = e\omega_J/\pi$.
The dashed lines show the effect of relatively large noise.
\label{Fig:IV-rsj-mod}}
\end{center}
\end{figure}

Then the simplified equation of motion reads
\begin{equation}
\omega_{Bc}^{-1}\, \dot{\theta}
+u_c \sin\theta\, [1+\mu \cos(\omega t + \phi_f)]
= v_0 +v_f.~~ \label{Eq-motion1-r2}
\end{equation}
This equation is similar to the RSJ model equation describing
an effective Josephson junction with a modulated critical current.
In the absence of modulation, $\mu=0$, and noise, $v_f=0$,
it takes the form
\begin{equation}
\omega_{c}^{-1}\, \dot{\theta}+\sin\theta =\nu, \label{mo-0}
\end{equation}
where
\begin{equation}
\omega_c=\omega_{Bc}u_c
= \pi V_c/eR_b~~~\textrm{and}~~~\nu =v_0/u_c. \label{omegaC}
\end{equation}
The solution of this equation reads \cite{KKLikharev-book}:
\begin{eqnarray}
\theta =2\arctan\left(\frac{\overline{i}}{\nu+1}\tan\frac{\Theta}{2}\right)
-\pi/2, \label{theta-t} \\
 \dot{\theta} = \overline{i}^2/(\nu-\sin \Theta),\label{i-t} \\
\overline{i}= (\nu^2- 1)^{1/2},   \label{iv-auto}   \\
\dot{\Theta}
= \omega_c \overline{i}(\nu),   \label{phase-leader}
\end{eqnarray}
where $\Theta$ is the phase leader related to the instantaneous
frequency of oscillations determined by
\begin{equation}
\overline{i}=\pi\overline{I}/e\omega_c=\omega_B/\omega_c. \label{IBnorm}
\end{equation}
Expression (\ref{iv-auto}) gives the autonomous IV-curve,
\begin{equation}
\overline{I_B}=(V_0^2-V_c^2)^{1/2}/R_b ,~~\label{I-V-hyperb}
\end{equation}
having standard hyperbolic shape, as shown in Fig.\,\ref{Fig:IV-rsj-mod}(a).
This curve is re-plotted in Fig.\,\ref{Fig:IV-rsj-mod}(b) as $\overline{I_B}$
against $\overline{V_B} = V_0 - R_b \overline{I_B}$.

%%%%%%%%%%%%%%%%%%%%%%%%%%%%%%%%%%%%%%%%%%%%%%%%%%%%
%%%%%%%%%%%%%%%%%%%%%%%%%%%%%%%%%%%%%%%%%%%%%%%%%%%%

\subsection{Formation of dual Shapiro step}%%%

In Josephson physics, Eq.\,(\ref{Eq-motion1-r2}) describes
the dynamics of a Josephson junction with an ac-modulated
critical current. This problem has been solved
by Vanneste \textit{et al.} in Ref.\,\cite{Vanneste-JLTP81},
where they showed that the I-V curve exhibited steps
resembling the conventional Shapiro steps.
Later on, such steps have been observed in the I-V curves
of a dc SQUID, modulated by an alternating magnetic flux
\cite{Vanneste-JAP88}, and in a recent
experiment \cite{Antonov2026}, as dual Shapiro steps,
when \emph{microwave drive} was applied to the BT island.
Below we show that these steps can appear
on the I-V curve of BT.

Assuming that parameter $\mu$ and the noise terms
in Eq.\,(\ref{Eq-motion1-r2})
are small, we will consider them as small perturbations.
In the vicinity of the expected step, the solution of
this equation can be presented in the form
\begin{equation}
\theta(t)=\theta_0(t) +\vartheta(t),~~\label{slow-var}
\end{equation}
where phase $\vartheta(t)$ is a slow variable,
i.e., $|\dot{\vartheta}|\ll\omega$.
Below, we will solve Eq.\,(\ref{Eq-motion1-r2}) using
the slowly varying phase method \cite{Bogoliubov-Mitropolski-book},
adopted by Likharev for an ac-driven Josephson junction
(see Chapter 10 in Ref.\,\cite{KKLikharev-book}).

We expand the needed solution $\theta(t)$ into the Taylor series
\begin{equation}
\theta = \theta_0+\vartheta_1+\vartheta_2+...,~~~~
\vartheta_k \propto \epsilon^k,~~~~k\geq 1, \label{theta-series}
\end{equation}
with respect to small $\epsilon \propto \mu,\,\phi_f,\,v_f$.
Then the corresponding "correction" to a dc voltage is
\begin{equation}
v_0 = \widehat{v}_0 + \widehat{v}_1 + \widehat{v}_2 +...,~~~~
\widehat{v}_k \propto \epsilon^k, \label{vb-series}
\end{equation}
where $\widehat{(...)}$ denotes "intermediate" averaging
over time interval $\Delta t$,
\begin{equation}
\omega^{-1}\ll \Delta t \ll |\Omega|^{-1}, \label{AverageTime}
\end{equation}
where $\Omega = \omega_B-\omega$ is the small detuning (beating) frequency.
Substitution of Eqs.\,(\ref{theta-series}) and (\ref{vb-series}) into        Eq.\,(\ref{Eq-motion1-r2}) yields the system of (the first three) equations:
\begin{equation}
\omega_{Bc}^{-1}\,\dot{\theta_0}
+u_c \sin\theta_0 = \widehat{v}_0,\label{theta0dyn}
\end{equation}
\begin{eqnarray}
\omega_{Bc}^{-1}\, \dot{\vartheta}_1 &+& u_c \cos\theta_0 \, \vartheta_1
 \nonumber \\
&=&-\mu u_c \sin \theta_0 \cos(\omega t + \phi_f)
+\widehat{v}_1 +v_f,  \label{theta1dyn}
\end{eqnarray}
\begin{equation}
\omega_{Bc}^{-1} \dot{\vartheta}_2 + u_c \cos\theta_0 \vartheta_2
= \widehat{v}_2 - 0.5 \sin \theta_0\, \vartheta_1^2.\label{theta2dyn}
\end{equation}

The solution of the first equation yields the unperturbed I-V
characteristic Eq.\,(\ref{I-V-hyperb})
and the frequency of autonomous oscillations
\begin{equation}
\omega_A = \pi I_A/e =\pi(V_0^2-V_c^2)^{1/2}/eR_b. \label{auto-omega}
\end{equation}
Following the approach developed in Ref.\,\cite{KKLikharev-book},
we find the voltage correction terms (\ref{vb-series}) using
the requirement of zero drift of phase, occurring due
to terms $\vartheta_1$, $\vartheta_2$,..., i.e.,
$\widehat{\dot{\vartheta_k}}=0$. Specifically, for $k=1$, we have
(cf. with Eq.\,(10.37) in Ref.\,\cite{KKLikharev-book})
\begin{equation}
[(\widehat{v}_1- v_{\mu}+v_f)(\widehat{v}_0 -\sin \Theta)]\widehat{~~} = 0,   ~~\label{corr-v1}
\end{equation}
where $\Theta$ is the phase leader.
Thus, we obtain the expression for $\widehat{v}_1$:
\begin{equation}
\widehat{v}_1 = \frac{\mu u_c}{2} \sin(\Theta - \omega t - \phi_f)
- \frac{( v_f \sin \Theta)\widehat{~~}}{v_0} .~~~\label{v1corr}
\end{equation}

Substituting $\widehat{v}_0+\widehat{v}_1$ (both are slow
variables) into the definition of the phase
leader Eq.\,(\ref{phase-leader}) and making the replacement
$\nu \rightarrow \nu + (v_1/u_c)$, we arrive at the equation
for $\Theta$:
\begin{equation}
\omega_c^{-1}\dot{\Theta} =  \overline{i}(\nu) + (d\overline{i}/d\nu)
v_1/u_c = \overline{i}_A + r^{-1}_d v_1(\Theta)/u_c,~~~\label{Theta-corr}
\end{equation}
where
$r_d^{-1} =\nu /\overline{i}=\sqrt{1+i^{-2}} \cong
\sqrt{1+(\omega_c/\omega)^2}$
is the dimensionless differential conductance
and $(r_d R_b)^{-1} =  d\overline{I}_B/dV_0$.
Here we expanded $\overline{i}(\nu)$ into the Taylor series and left
only the term $\propto (v_1/u_c)$. Thus, the explicit formula for
$\dot{\Theta}$ reads
\begin{equation}
\omega_c^{-1}\dot{\Theta}=\overline{i}+ r_d^{-1}
\left[\frac{\mu\sin(\Theta - \omega t - \phi_f)}{2}
-  \frac{(v_f \sin \Theta)\widehat{~~}}{u_cv_0}\right].\label{Theta-eq}
\end{equation}
Introducing a new (slow) variable, viz., the difference of the phase
leaders of the Bloch oscillations and of the
Josephson ac drive, $\Theta_r = \Theta - \omega t + \phi_f$, we obtain
the equation for $\Theta_r$:
\begin{eqnarray}
\dot{\Theta_r}=(\omega_A-\omega)
&+&0.5 r_d^{-1} \omega_c\mu\sin\Theta_r\nonumber\\
&+& \omega_{Jc} i_f-\omega_c(v_f\sin\omega t) \widehat{~~}.
\label{ThetaR-eq}
\end{eqnarray}
The first term on the right-hand side is the
difference between the frequency of autonomous oscillations
and the drive frequency $\omega$.
The second term ($\propto \sin \Theta_r$) ensures
phase locking of these two tones
and, hence, is responsible for the step formation.
In the absence of noise, the shape of this step mimics
the shape of the autonomous curve in the vicinity of the
Coulomb blockade ($\dot{\theta} = 0, I_B=0$).

The third term in Eq.\,(\ref{ThetaR-eq}) is related to
the noise, originated from the Josephson circuit.
This noise is transferred to the Bloch circuit due to
random drift $\phi_f$ of Josephson phase $\phi$
[see Eqs.\,(\ref{phi-linear}),(\ref{phi-f}), and (\ref{phi-Wiener})].
The fourth term is the noise originated from the Bloch circuit
itself and down-converted from the oscillation frequency $\omega$
in the same manner as the noise in an ac-driven Josephson
junction \cite{KKLikharev-book}.
Assuming that the noise sources
Eqs.\,(\ref{Sv-1}) and (\ref{SvSi}) are thermal with similar
temperature $T$ for both resistances $R_s$ and $R_b$
and, hence, are independent of frequency
($S_{I,V}\propto k_BT$, i.e., Johnson noise),
we can conveniently compare the contributions of the
third and the fourth terms:
\begin{equation}
\frac{\langle\omega_{Jc}i_f\rangle_{\textrm{rms}}}
{\langle\omega_c v_f\rangle_{\textrm{rms}}}
= 0.5A_{10} (\alpha_J/\alpha_B)^{1/2}\ll 1. \label{rms-ratio}
\end{equation}
Here we used Eqs.\,(\ref{omega-J}), (\ref{omegaC}), and
(\ref{V-threshold}), and strict inequality (\ref{alfaJ-alfaB}).
Thus we conclude that
the Bloch circuit noise, given by the fourth term
in Eq.\,(\ref{ThetaR-eq}), dominates.
Neglecting the third term makes Eq.\,(\ref{ThetaR-eq})
identical to that describing the effect of noise on the
conventional Shapiro step (see Ref.\,\cite{StephenPhysRev1969}
and  Chapter 10 in Ref.\,\cite{KKLikharev-book}).
The results of that study are obviously applicable to
our case. Then the dimensionless effective noise parameter is
\begin{equation}
\gamma_1 = 2 \Gamma_1^{(A)} (eR_d/\pi \mu V_c),
\label{gamma1}
\end{equation}
where $2\Gamma_1^{(A)}$ is the linewidth of the autonomous
Bloch oscillations.
For sufficiently high frequency, $\omega \gg \omega_c$,
the expression for the linewidth has a simple form,
$2\Gamma_1^{(A)}=2\pi e^{-2} k_BT/R_b$ (cf. Eq.\,(4.37)
in Ref.\,\cite{KKLikharev-book}).
The effect of relatively large noise,
$\gamma_1 \lesssim 1$, on the shape of the step is schematically
shown in the insets in Fig.\,\ref{Fig:IV-rsj-mod} by the dashed lines
(cf. Fig.\,10.5 in Ref.\,\cite{KKLikharev-book}).

\subsection{Size of the step}%%%

In essence, the obtained equation for the phase-leader
difference $\Theta_r$ (\ref{ThetaR-eq}) is the
RSJ model equation.
Neglecting noise, this equation has a stationary
solution, $\dot{\Theta}_r = 0$, existing in the range of
the autonomous oscillation frequency $\omega_A$ given by
the inequality
\begin{equation}
-\frac{\mu \omega_c}{2r_d} \leq \omega_A-\omega
\leq \frac{\mu \omega_c}{2r_d}.\label{synch-range}
\end{equation}
The resulting frequency range of phase-locking,
$\Delta \omega_A = \mu \omega_c/r_d$,
can be expressed in terms of the corresponding voltage $V_0$
[see Fig.\,\ref{Fig:IV-rsj-mod}(a)]:
\begin{equation}
\Delta V = r_d R_b (e/\pi) \Delta\omega_A
= \mu \frac{e\omega_c R_b}{\pi}
= \mu V_c. \label{synch-range-V}
\end{equation}
Thus, the step size is a factor of $\mu$ smaller than the
nominal blockade voltage $V_c$.

Interestingly, formula (\ref{synch-range-V}) is
valid for arbitrary $\mu$.
Neglecting the noise, solving general equation
of motion (\ref{Eq-motion1-r}) directly on the step (where Bloch
and Josephson frequencies are equal and have constant phase
difference $\Theta_r$)
is rather simple. Averaging both sides of this
equation over the period of oscillations, $2\pi/\omega
= 2\pi/\overline{\dot{\theta}}$, yields a dc term,
periodically dependent on phase difference $\Theta_r$.
Specifically, using the Fourier-series representation Eq.\,(\ref{v1v2}),
we have
\begin{equation}
\overline{v_B^{(0)}} = 0.5 \sum_{k\geq 1} k A_{kk}\sin(k \Theta_r)
\equiv F(\Theta_r).  \label{synch-range-V-general}
\end{equation}
Thus the step size,
\begin{equation}
\Delta V = \frac{\pi e}{2 C_\Sigma}
\left[ \max_{(0,\pi)}F(\Theta_r)
-\min_{(-\pi,0)} F(\Theta_r)\right],  \label{step-width-dV}
\end{equation}
scales as inverse capacitance $C_\Sigma^{-1}$ and depends on
the BT parameters $\lambda$ and $a$.

Taking into account the dominant (first) term in the infinite
series Eq.\,(\ref{synch-range-V-general}), we arrive at the
following relation:
\begin{equation}
\Delta V \approx \frac{\pi e A_{11}}{4 C_\Sigma}
[\sin(-\pi/2)-\sin(\pi/2)]
= \frac{\pi e}{2 C_\Sigma} |A_{11}|,  \label{step-width-dV1}
\end{equation}
or, using expression (\ref{V-threshold}) for blockade voltage
$V_c$,
\begin{equation}
\Delta V / V_c \approx 2|A_{11}|/A_{10} = \mu.  \label{step-dV-Vc}
\end{equation}
For example, for $\lambda=1$ and $a=0.1$, coefficients $A_{10}
\approx 0.5055$ and $A_{11} \approx -0.0979$ [see Eq.\,(\ref{matrix-A})]
give the relative step-size of $\Delta V/V_c \approx 0.39$.
Taking into account the first three terms in
Eq.\,(\ref{synch-range-V-general})
gives the value of $\Delta V/V_c \approx 0.42$.

\subsection{Transconductance}%%

%%%%%%%%%%%%%%%%%%%%%%%%%%%%%%%%%%%%%%%%%%%%%%%%%%%%
\begin{figure}%[b]
\begin{center}
\includegraphics[width=2.5in]{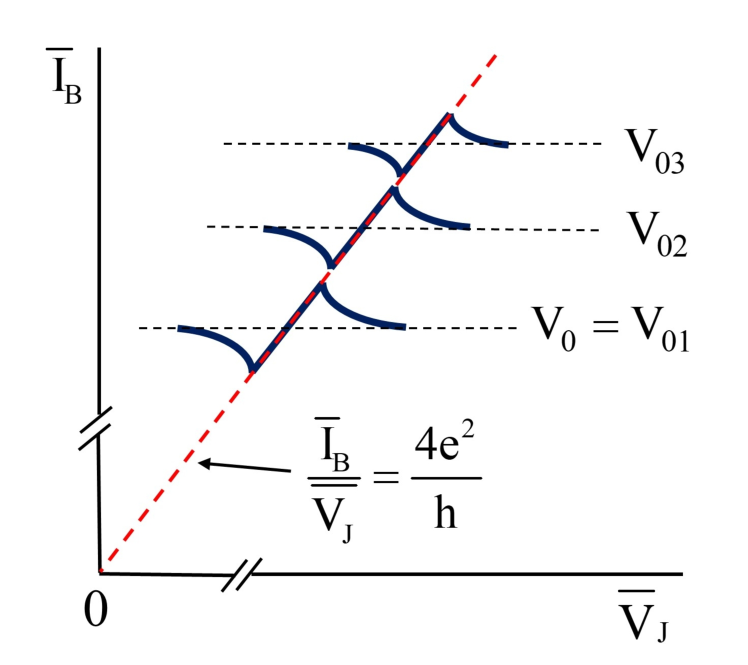}
\caption{Transconductance I-V curves exhibiting slanted
steps, where both the differential transconductance
$d\overline{I_B}/d\overline{V_J}$ and
the ratio $\overline{I_B}/\overline{V_J}$
take the fundamental value of $R^{-1}_Q$.
The curves are plotted for a sweep of current $I_0$
for three progressing fixed values of voltage $V_0$.
\label{Fig:transcond}}
\end{center}
\end{figure}
%%%%%%%%%%%%%%%%%%%%%%%%%%%%%%%%%%%%%%%%%%%%%%%%%%%%

When voltage $V_0$ is fixed such that average current
$\overline{I_A}$ and, hence, $\omega_A$ correspond
to the step center, the
synchronization range Eq.\,(\ref{synch-range}) can be
interpreted as the range of the Josephson
frequency $\omega = \omega_J$, i.e.,
\begin{equation}
\omega_A-\mu \omega_c/2r_d\leq\omega
\leq \omega_A+ \mu \omega_c/2r_d.\label{synch-range-omegaJ}
\end{equation}
Within this range the Bloch frequency, $\omega_B=
\dot{\Theta} = \pi \overline{I_B}/e$, is phase-locked by
the Josephson frequency, $\omega = (2\pi/\Phi_0)\overline{V_J}$,
i.e., $\omega_B=\omega$. So, within range
(\ref{synch-range-omegaJ}) the BT transconductance
is expressed via the fundamental constant, i.e.,
$\overline{I_B}/\overline{V_J} = R_Q^{-1}$.

The phase locking range can be expressed in terms of voltage
$\overline{V_J}$ as
\begin{equation}
\Delta \overline{V_J}= \frac{\Phi_0}{2\pi} \mu \omega_c/r_d
= \mu \alpha_B V_c/r_d,
\label{synch-range-VJ}
\end{equation}
or, in terms of current $\overline{I_B}$, as
\begin{equation}
\Delta \overline{I_B}= \Delta \overline{V_J}/R_Q = \mu V_c/R_d,
\label{synch-range-VJ}
\end{equation}
where the differential resistance is given by the RSJ model,
$R_d =r_d R_b=R_b/\sqrt{1+(V_c/R_b \overline{I_B})^2}$.

In the transconductance I-V characteristic, $\overline{I_B}$
versus $\overline{V_J}$, the phase-locking regime manifests
itself as a slanted step with the fundamental slope of $R_Q^{-1}$.
A series of such steps corresponding to different values of
voltage $V_0$ is shown in Fig.\,\ref{Fig:transcond}.
The red dashed line shows the locus of points
corresponding to synchronous steady motion of quasicharge $q$
and phase $\phi$, although with a velocity depending on the working
point fixed on this line. This behavior has a clear interpretation using
the egg box potential (see Fig.\,\ref{Fig:egg-box}), where
synchronous motion along the diagonal is obviously possible
at different tilt angles of the ground-energy plane.

%%%%%%%%%%%%%%%%%%%%%%%%%%%%%%%%%%%%%%%%%%%%%
\section{Discussion}%%%%%%%%%%%%%%%%%%%%%%%%%
%%%%%%%%%%%%%%%%%%%%%%%%%%%%%%%%%%%%%%%%%%%%%

We have described the mechanism of phase
locking of the Bloch and Josephson oscillations
assuming sufficiently high impedance of the Bloch circuit,
i.e., $\alpha_B =R_Q/R_b \ll 1$. This condition is crucially
important not only for our analysis Eq.\,(\ref{alfa-B}),
but also for the experimental observation of sufficiently
sharp steps. The reason is the noise associated with
the quantum uncertainty of the quasicharge because of its
coupling to the environment. Analysis of the case of not very
small $\alpha_B$ has been performed in Refs.\, \cite{DiMarco2015,Kurilovich2025,resch2025shapirosteps}.
The obtained dual Shapiro
steps have exhibited imperfect, smooth shapes. Specifically,
the accuracy of the step position was poor (see Fig.\,2 in
Ref.\,\cite{DiMarco2015}) and differential resistance
on the step was at best of the order of hundreds k$\Omega$
\cite{resch2025shapirosteps}. Thus, the strategy in designing
the experiment should include manufacturing sufficiently
high resistance $R_b$. Moreover, its parasitic capacitance
should contribute the smallest possible value to the total
capacitance of the island, $C_\Sigma$ \cite{Arndt2018}.
Then the values of the charging energy $E_c$, blockade
voltage $V_c$, and, hence, step size $\Delta V$ are not
notably diminished.

The step size $\Delta V$ Eq.\,(\ref{step-width-dV1})
plays a critical role in the experiment, when noise is
substantial. Large noise can arise, for example, because
of Joule heating of the bias resistance $R_b$, leading to the
increase of its effective electron
temperature, $T^* > T$ \cite{MaibaumPRB2011}.
Large $\Delta V$ forms a large energy barrier
$\Delta U=e\Delta V/2\pi$ preventing diffusion and jumps of phase
$\Theta_r$ and thereby ensures the flatness of the step.
For thermal (Markovian) noise, its effect on the step shape is
described by the corresponding Fokker-Plank equation
\cite{StephenPhysRev1969}.

\begin{figure}
\begin{center}
\includegraphics[width=2.6in]{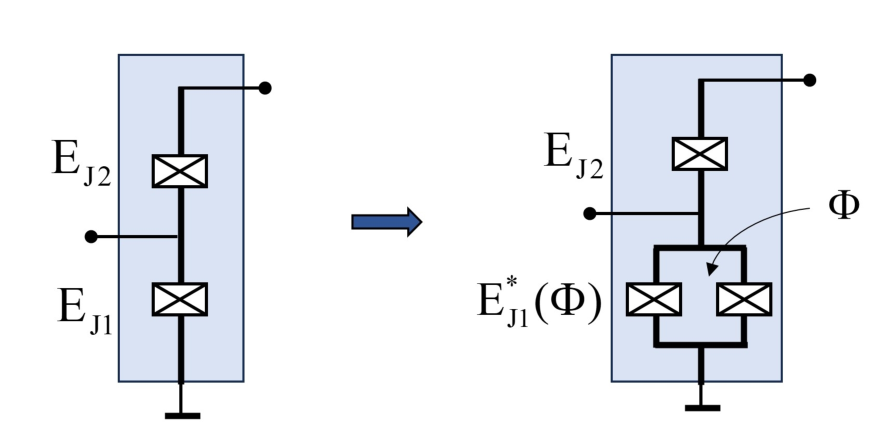}
\caption{Possible modification of BT enabling control of its
asymmetry, $r = E_{J2}/E^*_{J1}(\Phi)$, and, partially, the ratio
of characteristic energies $\lambda$
Eq.\,\textcolor{red}{(}\ref{EJEc-ratio}\textcolor{red}{)}.
The variation of effective critical current $I^*_{c1}$, and thereby
of the effective Josephson energy $E^*_{J1}$, is accomplished by
applying external magnetic flux $\Phi$ to the dc SQUID loop
encompassing two, nearly equal, small Josephson
junctions \cite{Clarke-Braginski-book}. Efficiency of controlling
the ratio of the Josephson and charging energies in a dc SQUID has
been experimentally demonstrated, e.g., in Ref.\,\cite{KaapNatComm2024}.
\label{Fig:BT-SQUID}}
\end{center}
\end{figure}

In the experiments with external microwave drive, the size
of a conventional Shapiro step (or a dual Shapiro step) is
optimized by adjusting the frequency and amplitude of the drive.
Although the amplitude of modulation $\mu$ cannot be
controlled in our circuit in the same
way as in experiments with external ac signal,
the obtained value of $\Delta V/V_c \approx 0.4$ is not small.
It is comparable with the size of the first Shapiro
step obtained in the RSJ model for an optimal amplitude of
ac drive \cite{LikharevSemenov1971} (see also
Fig.\,11.4 in Ref.\,\cite{KKLikharev-book}).
Interestingly, for achieving an appreciable size of the
step, the microwave frequency should be about $\omega_c$
that corresponds to a relatively large dc current on
the dual Shapiro step.
Such current causes notable power dissipation in the circuit
and, hence, elevated electron temperature $T^*$. The dc-driven Bloch
circuit can operate at somewhat lower frequency (and, hence,
lower current) without degradation of the step size.
Moreover, in our case, the step size can be further
optimized by adjusting the BT parameters. This can be done,
for example, using the BT design shown in
Fig.\,\ref{Fig:BT-SQUID}.
Embedding a small dc SQUID instead of a single Josephson
junction in one arm of the BT enables the control of the
BT symmetry, $r = E_{J2}/E^*_{J1}(\Phi)$, and, to some extent,
the control of the effective ratio of characteristic
energies $\lambda$. Thus the values of
matrix elements (\ref{matrix-A}) in this BT can be
tuned by external magnetic flux $\Phi$.
As a result, a somewhat larger (optimal) value of
coefficient $A_{11}$ in Eq.(\ref{step-width-dV1}) can
be achieved.

%%%%%%%%%%%%%%%%%%%%%%%%%%%%%%%%%%%%%%%%%%%%%%%%%%%%
%%%%%%%%%%%%%%%%%%%%%%%%%%%%%%%%%%%%%%%%%%%%%%%%%%%%

Furthermore, the absence of an external microwave
signal may mitigate the effect of noise in our system.
Specifically, depending on the circuit design, external
microwave irradiation may notably
enhance thermal noise in the circuit by means of heating
of the circuit elements.
This effect leads to broadening of both linewidths (primarily of the Bloch
oscillations) and, hence, further degradation of the step shape.
The telegraph $1/f$-like noise, associated mainly with random switching
of dipole-type two-level fluctuators, located in oxides around the BT island
\cite{AZ_Pashkin1998},
may also rise because of possible driving of these fluctuators out
of equilibrium.
However, for the-state-of-the-art circuits, exhibiting an
appreciable level of inherently wideband thermal noise
\cite{KaapNatComm2024, ShaikhaidarovNatCom2024},
the integral contribution of low-frequency telegraph
noise to the oscillation linewidth (see Eq.(4.33)
in Ref.\,\cite{KKLikharev-book})) is apparently small.
Thus the observation of dual Shapiro steps in
experiment with the \emph{quiet} BT circuit,
i.e., driven only by
dc sources, seems to be more favorable.

\begin{figure}
\begin{center}
\includegraphics[width=2.7in]{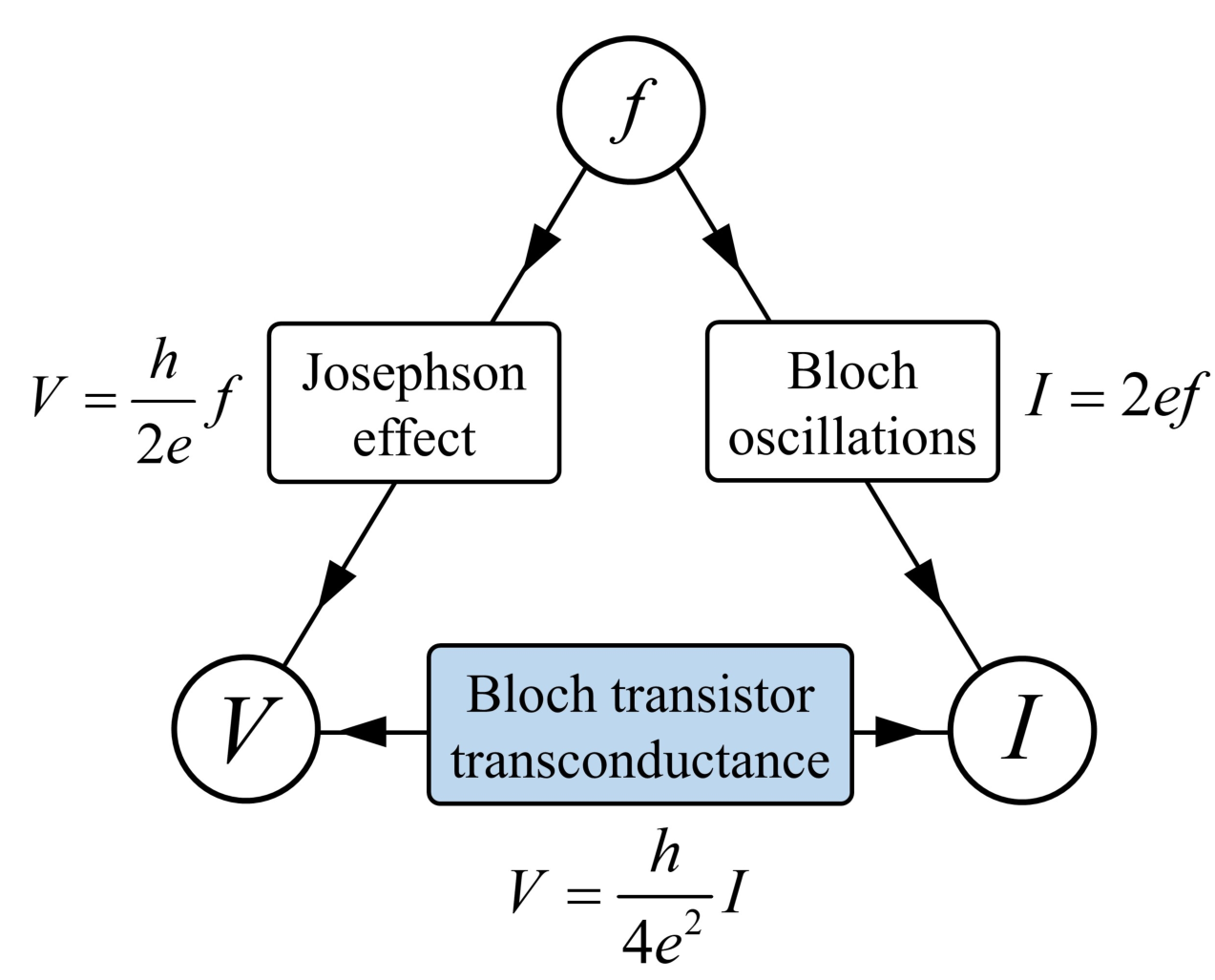}
\caption{Entirely \emph{superconducting} quantum
metrology triangle. Quantized Hall resistance, based
on a 2DEG in a high magnetic field, that was included in the
original version of the triangle for linking variable $V$
and $I$ \cite{KKL-AZ-JLTP85}, is here replaced by
the BT-based superconducting quantum resistance inserted
in the bottom side of the quantum metrology triangle.
\label{Fig:Triangle}}
\end{center}
\end{figure}

The obtained dual Shapiro steps are presumably horizontal,
i.e., having strictly constant current position.
The latter property is based on the assumption, that Josephson
frequency $\omega_J$ is fixed Eq.(\ref{phi-linear})
and, hence, independent of the working point on the step.
In fact, a change of the working point within the step causes
a small shift of the working point in the Josephson
circuit and, as a consequence, a small change of $\omega_J$.
In other words, it is a backaction of the Bloch circuit
on the Josephson circuit, that was not taken into account
in our model.
However, it is easy to show that in the case of sufficiently
low shunting resistance $R_s$ Eq.(\ref{alfaJ-alfaB}), this
effect is small and calculable. Specifically, the differential
resistance on the dual Shapiro step is equal to $R^2_Q/R_s$.
(This simple formula is derived in Appendix~B.)
For example, taking a rather conservative value of
$R_s = 1\,\Omega$, we obtain differential resistance
$R_d^{\textrm{(step)}}$ larger than $40\,\textrm{M}\Omega$.
Large value of $R_d^{\textrm{(step)}}$ allows considering
the effect of dual Shapiro steps in dc driven BT as a basis
for application in metrology. Moreover, the fact
that, first, the step center gives an exact value of current,
$\overline{I_B}=2ef_J$,
and, second, a very small step slope is well known, can
substantially weaken the effect of the circuit parameters.
Note that the slope of the transconductance step in
Fig.\,\ref{Fig:transcond} is independent of the
circuit parameters and, therefore, \emph{fundamental}.
This quantum phenomenon may complete the original quantum
metrology triangle \cite{KKL-AZ-JLTP85}
making it based entirely on superconducting components,
as shown in Fig.~\ref{Fig:Triangle}.

For the sake of brevity, we have focused only
on the fundamental dual Shapiro step, occurring
at $\omega_B = \omega_J$. Of course,
the BT properties allow coupling of harmonics of these frequencies,
i.e., $k\omega_B = m\omega_J$. This conclusion follows from the
premise that nonzero matrix element $A_{km}$ in Eq.(\ref{matrix-A})
naturally determines the size of the corresponding fractional step.
On the other hand, the absolute values of such matrix
elements are clearly smaller than $|A_{11}|$ and, consequently,
the experimental observation
of these fractional steps in the presence of appreciable
noise is more challenging.

%%%%%%%%%%%%%%%%%%%%%%%%%%%%%%%%%%%%%%%%%%%%%%%%%%%%
%%%%%%%%%%%%%%%%%%%%%%%%%%%%%%%%%%%%%%%%%%%%%%%%%%%%

%%%%%%%%%%%%%%%%%%%%%%%%%%%%%%%%%%%%%%%%%%%%
\section{Conclusion}%%%%%%%%%%%%%%%%%%%%%%%%%
%%%%%%%%%%%%%%%%%%%%%%%%%%%%%%%%%%%%%%%%%%%%%

In conclusion, we have proposed a circuit including BT
and two independent electrical sources controlling its
complementary variables, quasicharge $q$ and phase $\phi$.
We have shown that the shape of the ground state energy in BT
gives a unique possibility to mutually couple two different
types of oscillations, viz., the Josephson oscillations and Bloch
oscillations. There is a finite range of control parameters
where these oscillations are phase locked.
On the I-V curves of the Bloch circuit, this phase-locking
range manifests itself as a dual Shapiro step.
Realization of dual Shapiro steps using
remarkable properties of BT fed by dc sources could be an
interesting experiment per se.
The state of the art technology and proficiency, accumulated
in recent experiments, is a good basis for fast progress
in this experiment.

As long as the frequencies of Bloch and Josephson oscillations are
linked to the constant voltage and constant current by means of
the fundamental constants, $\Phi_0$ and $2e$, respectively, the
BT transconductance takes on the step the fundamental value of
$2e/\Phi_0=4e^2/h$.
The results of our study lead to the updated version of
the quantum metrology triangle, based entirely on superconducting
elements.
The proposed circuit paves the way to an alternative
quantum standard of resistance operating without a strong
magnetic field (cf. with the present 2DEG-based resistance
standard \cite{KvKlitzingRMP1986}).
After improvement of the experiment, this triangle
will certainly have a great impact
on possible refining of the values of $h$ and $e$, which
are the basis for modern electrical metrology \cite{Wiersma2021}.
Comparison of "superconducting" and 2DEG-based quantum
conductances with revising the value of the fine structure
constant could be another milestone goal.

We believe that our results will trigger further development
of the Bloch oscillation experiments based on advanced
fabrication technology, modern cryogenic instrumentation and
new ideas in designing superconducting quantum circuits.

\begin{acknowledgments}
The author is grateful to Yuri Pashkin for stimulating
discussions and useful comments.
\end{acknowledgments}

\section*{data availability}
The data that support the findings of this article are openly
available \cite{Supp_Info_Zenodo2026}.

\appendix

\section{Fourier series representation of $V_B(q,\phi)$ and $I_s(q,\phi)$}

Due to the double periodicity Eq.(\ref{En-periodicity}) and evenness Eq.(\ref{En-symmetry}) of the ground-state energy
$E_0(q,\phi)$, it can be represented as a double Fourier series,
\begin{eqnarray}
\varepsilon(\theta,\phi) &=& -0.5\sum_{k\geq 1} A_{k0}\cos(k \theta)
-0.5\sum_{m\geq 1} A_{0m} \cos (m \phi)   \nonumber \\
&-& \sum_{k,m\geq 1} A_{km} \cos(k \theta)\cos (m \phi) - 0.25A_{00}, \label{E0-series}
\end{eqnarray}
where the dimensionless energy and quasicharge are
\begin{equation}
\varepsilon = E_0(q,\phi)/E_c~~\textrm{and}~~ \theta = \pi q/e. \label{E0-norm}
\end{equation}
The real Fourier coefficients in Eq.(\ref{E0-series})  are
\begin{equation}
A_{km} = -\frac{4}{\pi^2} \int_{0}^{\pi}\int_{0}^{\pi}
\varepsilon(\theta,\phi)
\cos(k \theta)\cos (m \phi)d\theta d\phi,~~ \label{FourierCoeff}
\end{equation}
where $k$ and $m$ are non-negative integer indexes,
$k\geq 0$ and $m\geq 0$.

Elements $A_{km}$ of infinite matrix $\textbf{A}$ depend on
the energy ratio parameter $\lambda$ Eq.(\ref{EJEc-ratio})
and the asymmetry factor $a$ Eq.(\ref{asymm}).
For typical experimental values of BT parameters
($\lambda \sim 1$ and not extremely small $a$), the ground
energy profile $E_0(q,\phi)$ is obviously smooth and
the absolute values $|A_{km}|$ steeply decay with a rise
of $k$ or/and $m$. As an illustration, the most essential
part of matrix $\textbf{A}$,
numerically calculated for $\lambda = 1$ and $a=0.1$,
is given below \cite{Supp_Info_Zenodo2026}:
\begin{equation}
\textbf{A} =
\begin{bmatrix}
& ~~0.6037 & -0.0386 &  ~~0.0094 & ...\\
~~\textbf{0.5055}&-\textbf{0.0979}& ~~0.0237&-0.0064&...\\
-0.0685	& ~~0.0317& -0.0126	& ~~0.0051 & ...\\
~~0.0193 & -0.0112 & ~~0.0067&-0.0034 & ...\\
 ... & ... & ... & ...&...
\end{bmatrix}
, \label{matrix-A}
\end{equation}
where unimportant matrix element $A_{00}$, associated with
an additive constant for energy $\varepsilon$, is omitted.
The dominant matrix elements,
$A_{10}$ and $A_{11}$, giving approximate
width of the dual Shapiro step in Eq.\,(\ref{step-dV-Vc})
are marked in bold font.

Using Eqs.(\ref{voltage-dEdq}) and (\ref{Is-via-Q2-2})
we obtain the Fourier series for dimensionless voltage
$v_B^{(0)}=V_B^{(0)}/V_u$ and current $i_s=I_s/I_u$, i.e.,
\begin{eqnarray}
v_B^{(0)}(\theta,\phi) &=& 0.5\sum_{k\geq 1}k A_{k0}\sin(k \theta) \nonumber \\
&+& \sum_{k,m\geq 1}k A_{km}\sin(k \vartheta)\cos (m \phi) \label{v1v2}
\end{eqnarray}
 and
\begin{eqnarray}
i_s(\theta,\phi) &=& 0.5\sum_{m\geq 1}m A_{0m}\sin(m \phi) \nonumber \\
&+& \sum_{k,m\geq 1} m A_{km} \cos(k \theta)\sin (m \phi), \label{i1i2}
\end{eqnarray}
respectively.
The first sum in Eq.\,(\ref{v1v2}) [Eq.\,(\ref{i1i2})]
is independent of variable $\phi$ (variable $\theta$).
So, this part of the voltage (current) is decoupled
from the complementary variable phase $\phi$
(quasicharge $\theta$). The first terms in these series
dominate and, hence, they give the estimates of the
blockade voltage
and the critical current of BT, respectively.
Specifically, voltage
\begin{equation}
V_c \approx \frac{\pi e}{4 C_\Sigma} A_{10};
\label{V-threshold}
\end{equation}
similarly, the critical current is
\begin{equation}
I_c \sim \frac{\pi e}{4 R_Q C_\Sigma} A_{01}.
\label{Ic-approx}
\end{equation}
Because the current-phase relation in BT is inherently
nonsinusoidal \cite{AZ-PRL1996}, the latter relation,
containing only one term from the entire double
series (\ref{i1i2}), is a rough estimate.

\section{Slope of dual Shapiro step}

To find the correction to Josephson frequency $\omega_J$,
we take into account relatively small term $i_s$ in
Eq.\,(\ref{Eq-motion2Norm}). On the step, the slowly varying part
of current $i_s$ is given by formula (\ref{i1i2}),
\begin{equation}
\widehat{i}_s =-0.5\sum_{m\geq 1}m A_{mm}\sin(m \Theta_r)
\approx -0.5 A_{11}\sin\Theta_r.
\label{is-dc}
\end{equation}
As a result, the constant current term $i_0$ in
Eq.\,(\ref{Eq-motion2Norm}) takes the form,
\begin{equation}
i_0 \rightarrow i_0+\Delta i_0,~~\textrm{where}~~\Delta i_0
= 0.5 A_{11}\sin\Theta_r.
\label{i0-dc-corr}
\end{equation}
The corresponding correction to dc voltage $\overline{V_J}$ is
\begin{equation}
\Delta \overline{V_J} = \Delta i_0 I_u R^{(J)}_d \approx
\alpha_J \frac{\pi e}{2C_\Sigma} \Delta i_0,
\label{dV_J}
\end{equation}
where we used the fact that the differential resistance in the Josephson
circuit is $R^{(J)}_d \approx R_s$.

On the step, the Josephson and
Bloch frequencies are equal and the difference of their phases
$\Theta_r$ is fixed. Hence, the above correction to
$\overline{V_J}$ is straightforwardly translated to corrections
of $\overline{I_B}$, i.e.,
\begin{equation}
\Delta \overline{I_B}(\Theta_r) = R^{-1}_Q \Delta \overline{V_J}(\Theta_r).
\label{dI_B}
\end{equation}
Thus the total change of current $\overline{I_B}$ at scanning
from one edge of the step ($\Theta_r=-\pi/2$) to the
other ($\Theta_r=\pi/2$) is
\begin{equation}
\delta \overline{I_B}= \Delta \overline{I_B}(-\pi/2)
- \Delta \overline{I_B}(\pi/2)
= \frac{\pi e \alpha_J}{2R_Q C_\Sigma}|A_{11}|.
\label{delta-I_B}
\end{equation}
Comparison of Eq.\,(\ref{delta-I_B}) with formula (\ref{step-width-dV1})
for the dual Shapiro step size, yields the differential
resistance on the step, i.e.,
\begin{equation}
R^{(\textrm{step})}_d = \Delta V/\delta \overline{I_B}
= R_Q/\alpha_J = R^2_Q/R_s.
\label{conduct-step}
\end{equation}


\begin{thebibliography}{65}%
\makeatletter
\providecommand \@ifxundefined [1]{%
 \@ifx{#1\undefined}
}%
\providecommand \@ifnum [1]{%
 \ifnum #1\expandafter \@firstoftwo
 \else \expandafter \@secondoftwo
 \fi
}%
\providecommand \@ifx [1]{%
 \ifx #1\expandafter \@firstoftwo
 \else \expandafter \@secondoftwo
 \fi
}%
\providecommand \natexlab [1]{#1}%
\providecommand \enquote  [1]{``#1''}%
\providecommand \bibnamefont  [1]{#1}%
\providecommand \bibfnamefont [1]{#1}%
\providecommand \citenamefont [1]{#1}%
\providecommand \href@noop [0]{\@secondoftwo}%
\providecommand \href [0]{\begingroup \@sanitize@url \@href}%
\providecommand \@href[1]{\@@startlink{#1}\@@href}%
\providecommand \@@href[1]{\endgroup#1\@@endlink}%
\providecommand \@sanitize@url [0]{\catcode `\\12\catcode `\$12\catcode
  `\&12\catcode `\#12\catcode `\^12\catcode `\_12\catcode `\%12\relax}%
\providecommand \@@startlink[1]{}%
\providecommand \@@endlink[0]{}%
\providecommand \url  [0]{\begingroup\@sanitize@url \@url }%
\providecommand \@url [1]{\endgroup\@href {#1}{\urlprefix }}%
\providecommand \urlprefix  [0]{URL }%
\providecommand \Eprint [0]{\href }%
\providecommand \doibase [0]{https://doi.org/}%
\providecommand \selectlanguage [0]{\@gobble}%
\providecommand \bibinfo  [0]{\@secondoftwo}%
\providecommand \bibfield  [0]{\@secondoftwo}%
\providecommand \translation [1]{[#1]}%
\providecommand \BibitemOpen [0]{}%
\providecommand \bibitemStop [0]{}%
\providecommand \bibitemNoStop [0]{.\EOS\space}%
\providecommand \EOS [0]{\spacefactor3000\relax}%
\providecommand \BibitemShut  [1]{\csname bibitem#1\endcsname}%
\let\auto@bib@innerbib\@empty
%</preamble>
\bibitem [{\citenamefont {Pashkin}\ \emph {et~al.}(2009)\citenamefont
  {Pashkin}, \citenamefont {Astafiev}, \citenamefont {Yamamoto}, \citenamefont
  {Nakamura},\ and\ \citenamefont {Tsai}}]{Pashkin2009}%
  \BibitemOpen
  \bibfield  {author} {\bibinfo {author} {\bibfnamefont {Y.~A.}\ \bibnamefont
  {Pashkin}}, \bibinfo {author} {\bibfnamefont {O.}~\bibnamefont {Astafiev}},
  \bibinfo {author} {\bibfnamefont {T.}~\bibnamefont {Yamamoto}}, \bibinfo
  {author} {\bibfnamefont {Y.}~\bibnamefont {Nakamura}},\ and\ \bibinfo
  {author} {\bibfnamefont {J.~S.}\ \bibnamefont {Tsai}},\ }\bibfield  {title}
  {\bibinfo {title} {Josephson charge qubits: a brief review},\ }\href
  {https://doi.org/https://doi.org/10.1007/s11128-009-0101-5} {\bibfield
  {journal} {\bibinfo  {journal} {Quantum Inf. Processing}\ }\textbf {\bibinfo
  {volume} {8}},\ \bibinfo {pages} {55} (\bibinfo {year} {2009})}\BibitemShut
  {NoStop}%
\bibitem [{\citenamefont {Koch}\ \emph {et~al.}(2007)\citenamefont {Koch},
  \citenamefont {Yu}, \citenamefont {Gambetta}, \citenamefont {Houck},
  \citenamefont {Schuster}, \citenamefont {Majer}, \citenamefont {Blais},
  \citenamefont {Devoret}, \citenamefont {Girvin},\ and\ \citenamefont
  {Schoelkopf}}]{Koch2007}%
  \BibitemOpen
  \bibfield  {author} {\bibinfo {author} {\bibfnamefont {J.}~\bibnamefont
  {Koch}}, \bibinfo {author} {\bibfnamefont {T.~M.}\ \bibnamefont {Yu}},
  \bibinfo {author} {\bibfnamefont {J.}~\bibnamefont {Gambetta}}, \bibinfo
  {author} {\bibfnamefont {A.~A.}\ \bibnamefont {Houck}}, \bibinfo {author}
  {\bibfnamefont {D.~I.}\ \bibnamefont {Schuster}}, \bibinfo {author}
  {\bibfnamefont {J.}~\bibnamefont {Majer}}, \bibinfo {author} {\bibfnamefont
  {A.}~\bibnamefont {Blais}}, \bibinfo {author} {\bibfnamefont {M.~H.}\
  \bibnamefont {Devoret}}, \bibinfo {author} {\bibfnamefont {S.~M.}\
  \bibnamefont {Girvin}},\ and\ \bibinfo {author} {\bibfnamefont {R.~J.}\
  \bibnamefont {Schoelkopf}},\ }\bibfield  {title} {\bibinfo {title}
  {Charge-insensitive qubit design derived from the {C}ooper pair box},\ }\href
  {https://doi.org/10.1103/PhysRevA.76.042319} {\bibfield  {journal} {\bibinfo
  {journal} {Phys. Rev. A}\ }\textbf {\bibinfo {volume} {76}},\ \bibinfo
  {pages} {042319} (\bibinfo {year} {2007})}\BibitemShut {NoStop}%
\bibitem [{\citenamefont {Clarke}\ and\ \citenamefont
  {Wilhelm}(2008)}]{Clarke2008}%
  \BibitemOpen
  \bibfield  {author} {\bibinfo {author} {\bibfnamefont {J.}~\bibnamefont
  {Clarke}}\ and\ \bibinfo {author} {\bibfnamefont {F.}~\bibnamefont
  {Wilhelm}},\ }\bibfield  {title} {\bibinfo {title} {Superconducting quantum
  bits},\ }\href {https://doi.org/https://doi.org/10.1038/nature07128}
  {\bibfield  {journal} {\bibinfo  {journal} {Nature}\ }\textbf {\bibinfo
  {volume} {453}},\ \bibinfo {pages} {1031} (\bibinfo {year}
  {2008})}\BibitemShut {NoStop}%
\bibitem [{\citenamefont {Manucharyan}\ \emph {et~al.}(2009)\citenamefont
  {Manucharyan}, \citenamefont {Koch}, \citenamefont {Glazman},\ and\
  \citenamefont {Devoret}}]{Manucharyan2009}%
  \BibitemOpen
  \bibfield  {author} {\bibinfo {author} {\bibfnamefont {V.~E.}\ \bibnamefont
  {Manucharyan}}, \bibinfo {author} {\bibfnamefont {J.}~\bibnamefont {Koch}},
  \bibinfo {author} {\bibfnamefont {L.~I.}\ \bibnamefont {Glazman}},\ and\
  \bibinfo {author} {\bibfnamefont {M.~H.}\ \bibnamefont {Devoret}},\
  }\bibfield  {title} {\bibinfo {title} {Fluxonium: {S}ingle {C}ooper-pair
  circuit free of charge offsets},\ }\href
  {https://doi.org/https://doi.org/10.1126/science.1175552} {\bibfield
  {journal} {\bibinfo  {journal} {Science}\ }\textbf {\bibinfo {volume}
  {326}},\ \bibinfo {pages} {113} (\bibinfo {year} {2009})}\BibitemShut
  {NoStop}%
\bibitem [{\citenamefont {Makhlin}\ \emph {et~al.}(2001)\citenamefont
  {Makhlin}, \citenamefont {Sch\"on},\ and\ \citenamefont
  {Shnirman}}]{MakhlinRMP2001}%
  \BibitemOpen
  \bibfield  {author} {\bibinfo {author} {\bibfnamefont {Y.}~\bibnamefont
  {Makhlin}}, \bibinfo {author} {\bibfnamefont {G.}~\bibnamefont {Sch\"on}},\
  and\ \bibinfo {author} {\bibfnamefont {A.}~\bibnamefont {Shnirman}},\
  }\bibfield  {title} {\bibinfo {title} {Quantum-state engineering with
  {J}osephson-junction devices},\ }\href
  {https://doi.org/10.1103/RevModPhys.73.357} {\bibfield  {journal} {\bibinfo
  {journal} {Rev. Mod. Phys.}\ }\textbf {\bibinfo {volume} {73}},\ \bibinfo
  {pages} {357} (\bibinfo {year} {2001})}\BibitemShut {NoStop}%
\bibitem [{\citenamefont {Pechenezhskiy}\ \emph {et~al.}(2020)\citenamefont
  {Pechenezhskiy}, \citenamefont {Mencia}, \citenamefont {Nguyen},
  \citenamefont {Lin},\ and\ \citenamefont {Manucharyan}}]{Pechenezhskiy2020}%
  \BibitemOpen
  \bibfield  {author} {\bibinfo {author} {\bibfnamefont {I.~V.}\ \bibnamefont
  {Pechenezhskiy}}, \bibinfo {author} {\bibfnamefont {R.~A.}\ \bibnamefont
  {Mencia}}, \bibinfo {author} {\bibfnamefont {L.~B.}\ \bibnamefont {Nguyen}},
  \bibinfo {author} {\bibfnamefont {Y.~H.}\ \bibnamefont {Lin}},\ and\ \bibinfo
  {author} {\bibfnamefont {V.~E.}\ \bibnamefont {Manucharyan}},\ }\bibfield
  {title} {\bibinfo {title} {The superconducting quasicharge qubit},\ }\href
  {https://doi.org/DOI: 10.1038/s41586-020-2687-9} {\bibfield  {journal}
  {\bibinfo  {journal} {Nature}\ }\textbf {\bibinfo {volume} {585}},\ \bibinfo
  {pages} {368} (\bibinfo {year} {2020})}\BibitemShut {NoStop}%
\bibitem [{\citenamefont {Averin}\ \emph {et~al.}(1985)\citenamefont {Averin},
  \citenamefont {Zorin},\ and\ \citenamefont {Likharev}}]{DA-AZ-KKL-JETP1985}%
  \BibitemOpen
  \bibfield  {author} {\bibinfo {author} {\bibfnamefont {D.~V.}\ \bibnamefont
  {Averin}}, \bibinfo {author} {\bibfnamefont {A.~B.}\ \bibnamefont {Zorin}},\
  and\ \bibinfo {author} {\bibfnamefont {K.~K.}\ \bibnamefont {Likharev}},\
  }\bibfield  {title} {\bibinfo {title} {Bloch oscillations in small
  {J}osephson junctions},\ }\href
  {https://www.jetp.ras.ru/cgi-bin/dn/e_061_02_0407.pdf} {\bibfield  {journal}
  {\bibinfo  {journal} {Zh. Eksp. Teor. Fiz.}\ }\textbf {\bibinfo {volume}
  {88}},\ \bibinfo {pages} {692} (\bibinfo {year} {1985})},\ \bibinfo {note}
  {[Sov. Phys. JETP \textbf{61}, 407-413 (1985)]}\BibitemShut {NoStop}%
\bibitem [{\citenamefont {Likharev}\ and\ \citenamefont
  {Zorin}(1985)}]{KKL-AZ-JLTP85}%
  \BibitemOpen
  \bibfield  {author} {\bibinfo {author} {\bibfnamefont {K.~K.}\ \bibnamefont
  {Likharev}}\ and\ \bibinfo {author} {\bibfnamefont {A.~B.}\ \bibnamefont
  {Zorin}},\ }\bibfield  {title} {\bibinfo {title} {Theory of the {B}loch-wave
  oscillations in small {J}osephson junctions},\ }\href
  {https://doi.org/https://doi.org/10.1007/BF00683782} {\bibfield  {journal}
  {\bibinfo  {journal} {J. Low Temp. Phys.}\ }\textbf {\bibinfo {volume}
  {59}},\ \bibinfo {pages} {347–382} (\bibinfo {year} {1985})}\BibitemShut
  {NoStop}%
\bibitem [{\citenamefont {Josephson}(1962)}]{JOSEPHSON1962}%
  \BibitemOpen
  \bibfield  {author} {\bibinfo {author} {\bibfnamefont {B.~D.}\ \bibnamefont
  {Josephson}},\ }\bibfield  {title} {\bibinfo {title} {Possible new effects in
  superconductive tunnelling},\ }\href
  {https://doi.org/https://doi.org/10.1016/0031-9163(62)91369-0} {\bibfield
  {journal} {\bibinfo  {journal} {Physics Letters}\ }\textbf {\bibinfo {volume}
  {1}},\ \bibinfo {pages} {251} (\bibinfo {year} {1962})}\BibitemShut {NoStop}%
\bibitem [{\citenamefont {Kuzmin}\ and\ \citenamefont
  {Haviland}(1991)}]{KuzminHaviland1991}%
  \BibitemOpen
  \bibfield  {author} {\bibinfo {author} {\bibfnamefont {L.~S.}\ \bibnamefont
  {Kuzmin}}\ and\ \bibinfo {author} {\bibfnamefont {D.~B.}\ \bibnamefont
  {Haviland}},\ }\bibfield  {title} {\bibinfo {title} {Observation of the
  {B}loch oscillations in an ultrasmall {J}osephson junction},\ }\href
  {https://doi.org/10.1103/PhysRevLett.67.2890} {\bibfield  {journal} {\bibinfo
   {journal} {Phys. Rev. Lett.}\ }\textbf {\bibinfo {volume} {67}},\ \bibinfo
  {pages} {2890} (\bibinfo {year} {1991})}\BibitemShut {NoStop}%
\bibitem [{\citenamefont {Kuzmin}\ and\ \citenamefont
  {Haviland}(1992)}]{KuzminHaviland1992}%
  \BibitemOpen
  \bibfield  {author} {\bibinfo {author} {\bibfnamefont {L.~S.}\ \bibnamefont
  {Kuzmin}}\ and\ \bibinfo {author} {\bibfnamefont {D.~B.}\ \bibnamefont
  {Haviland}},\ }\bibfield  {title} {\bibinfo {title} {Bloch oscillations and
  {C}oulomb blockade of {C}ooper pair tunneling in ultrasmall {J}osephson
  junctions},\ }\href {https://doi.org/10.1088/0031-8949/1992/T42/029}
  {\bibfield  {journal} {\bibinfo  {journal} {Phys. Scripta}\ }\textbf
  {\bibinfo {volume} {T42}},\ \bibinfo {pages} {171} (\bibinfo {year}
  {1992})}\BibitemShut {NoStop}%
\bibitem [{\citenamefont {Kuzmin}\ \emph {et~al.}(1994)\citenamefont {Kuzmin},
  \citenamefont {Pashkin},\ and\ \citenamefont
  {Claeson}}]{KuzminPashkinClaeson1994}%
  \BibitemOpen
  \bibfield  {author} {\bibinfo {author} {\bibfnamefont {L.~S.}\ \bibnamefont
  {Kuzmin}}, \bibinfo {author} {\bibfnamefont {Y.~A.}\ \bibnamefont
  {Pashkin}},\ and\ \bibinfo {author} {\bibfnamefont {T.}~\bibnamefont
  {Claeson}},\ }\bibfield  {title} {\bibinfo {title} {Bloch oscillations in a
  double {J}osephson junction biased via high-ohmic resistors},\ }\href
  {https://doi.org/10.1088/0953-2048/7/5/025} {\bibfield  {journal} {\bibinfo
  {journal} {Supercond. Sci. Technol.}\ }\textbf {\bibinfo {volume} {7}},\
  \bibinfo {pages} {324} (\bibinfo {year} {1994})}\BibitemShut {NoStop}%
\bibitem [{\citenamefont {Pashkin}\ \emph {et~al.}(1996)\citenamefont
  {Pashkin}, \citenamefont {Chen}, \citenamefont {Haviland},\ and\
  \citenamefont {Kuzmin}}]{Pashkin1996}%
  \BibitemOpen
  \bibfield  {author} {\bibinfo {author} {\bibfnamefont {Y.~A.}\ \bibnamefont
  {Pashkin}}, \bibinfo {author} {\bibfnamefont {C.~D.}\ \bibnamefont {Chen}},
  \bibinfo {author} {\bibfnamefont {D.~B.}\ \bibnamefont {Haviland}},\ and\
  \bibinfo {author} {\bibfnamefont {L.~S.}\ \bibnamefont {Kuzmin}},\ }\bibfield
   {title} {\bibinfo {title} {Magnetic field dependence of the current-voltage
  curve of a superconducting single electron transistor in a high impedance
  environment},\ }\href {https://doi.org/https://doi.org/10.1007/BF02571137}
  {\bibfield  {journal} {\bibinfo  {journal} {Czech. J. Phys.}\ }\textbf
  {\bibinfo {volume} {46}},\ \bibinfo {pages} {2291} (\bibinfo {year}
  {1996})}\BibitemShut {NoStop}%
\bibitem [{\citenamefont {Watanabe}\ and\ \citenamefont
  {Haviland}(2001)}]{WatanabeHaviland2001}%
  \BibitemOpen
  \bibfield  {author} {\bibinfo {author} {\bibfnamefont {M.}~\bibnamefont
  {Watanabe}}\ and\ \bibinfo {author} {\bibfnamefont {D.~B.}\ \bibnamefont
  {Haviland}},\ }\bibfield  {title} {\bibinfo {title} {Coulomb blockade and
  coherent single-{C}ooper-pair tunneling in single {J}osephson junctions},\
  }\href {https://doi.org/10.1103/PhysRevLett.86.5120} {\bibfield  {journal}
  {\bibinfo  {journal} {Phys. Rev. Lett.}\ }\textbf {\bibinfo {volume} {86}},\
  \bibinfo {pages} {5120} (\bibinfo {year} {2001})}\BibitemShut {NoStop}%
\bibitem [{\citenamefont {Lotkhov}(2013)}]{Lotkhov2013}%
  \BibitemOpen
  \bibfield  {author} {\bibinfo {author} {\bibfnamefont {S.~V.}\ \bibnamefont
  {Lotkhov}},\ }\bibfield  {title} {\bibinfo {title} {Ultra-high-ohmic
  microstripline resistors for {C}oulomb blockade devices},\ }\href
  {https://doi.org/10.1088/0957-4484/24/23/235201} {\bibfield  {journal}
  {\bibinfo  {journal} {Nanotechnology}\ }\textbf {\bibinfo {volume} {24}},\
  \bibinfo {pages} {235201} (\bibinfo {year} {2013})}\BibitemShut {NoStop}%
\bibitem [{\citenamefont {Gr\"unhaupt}\ \emph {et~al.}(2019)\citenamefont
  {Gr\"unhaupt}, \citenamefont {Spiecker}, \citenamefont {Gusenkova},
  \citenamefont {Maleeva}, \citenamefont {Skacel}, \citenamefont {Takmakov},
  \citenamefont {Valenti}, \citenamefont {Winkel}, \citenamefont {Rotzinger},
  \citenamefont {Wernsdorfer}, \citenamefont {Ustinov},\ and\ \citenamefont
  {Pop}}]{Grunhaupt2019}%
  \BibitemOpen
  \bibfield  {author} {\bibinfo {author} {\bibfnamefont {L.}~\bibnamefont
  {Gr\"unhaupt}}, \bibinfo {author} {\bibfnamefont {M.}~\bibnamefont
  {Spiecker}}, \bibinfo {author} {\bibfnamefont {D.}~\bibnamefont {Gusenkova}},
  \bibinfo {author} {\bibfnamefont {N.}~\bibnamefont {Maleeva}}, \bibinfo
  {author} {\bibfnamefont {S.~T.}\ \bibnamefont {Skacel}}, \bibinfo {author}
  {\bibfnamefont {I.}~\bibnamefont {Takmakov}}, \bibinfo {author}
  {\bibfnamefont {F.}~\bibnamefont {Valenti}}, \bibinfo {author} {\bibfnamefont
  {P.}~\bibnamefont {Winkel}}, \bibinfo {author} {\bibfnamefont
  {H.}~\bibnamefont {Rotzinger}}, \bibinfo {author} {\bibfnamefont
  {W.}~\bibnamefont {Wernsdorfer}}, \bibinfo {author} {\bibfnamefont {A.~V.}\
  \bibnamefont {Ustinov}},\ and\ \bibinfo {author} {\bibfnamefont {I.~M.}\
  \bibnamefont {Pop}},\ }\bibfield  {title} {\bibinfo {title} {Granular
  aluminium as a superconducting material for high-impedance quantum
  circuits},\ }\href
  {https://doi.org/https://doi.org/10.1038/s41563-019-0350-3} {\bibfield
  {journal} {\bibinfo  {journal} {Nat. Materials}\ }\textbf {\bibinfo {volume}
  {18}},\ \bibinfo {pages} {816} (\bibinfo {year} {2019})}\BibitemShut
  {NoStop}%
\bibitem [{\citenamefont {Kaap}\ \emph
  {et~al.}(2024{\natexlab{a}})\citenamefont {Kaap}, \citenamefont {Scheer},
  \citenamefont {Hassler},\ and\ \citenamefont {Lotkhov}}]{KaapPRL2024}%
  \BibitemOpen
  \bibfield  {author} {\bibinfo {author} {\bibfnamefont {F.}~\bibnamefont
  {Kaap}}, \bibinfo {author} {\bibfnamefont {D.}~\bibnamefont {Scheer}},
  \bibinfo {author} {\bibfnamefont {F.}~\bibnamefont {Hassler}},\ and\ \bibinfo
  {author} {\bibfnamefont {S.}~\bibnamefont {Lotkhov}},\ }\bibfield  {title}
  {\bibinfo {title} {Synchronization of {B}loch oscillations in a strongly
  coupled pair of small {J}osephson junctions: Evidence for a {S}hapiro-like
  current step},\ }\href {https://doi.org/10.1103/PhysRevLett.132.027001}
  {\bibfield  {journal} {\bibinfo  {journal} {Phys. Rev. Lett.}\ }\textbf
  {\bibinfo {volume} {132}},\ \bibinfo {pages} {027001} (\bibinfo {year}
  {2024}{\natexlab{a}})}\BibitemShut {NoStop}%
\bibitem [{\citenamefont {Kaap}\ \emph
  {et~al.}(2024{\natexlab{b}})\citenamefont {Kaap}, \citenamefont {Kissling},
  \citenamefont {Gaydamachenko}, \citenamefont {Gr\"{u}nhaupt},\ and\
  \citenamefont {Lotkhov}}]{KaapNatComm2024}%
  \BibitemOpen
  \bibfield  {author} {\bibinfo {author} {\bibfnamefont {F.}~\bibnamefont
  {Kaap}}, \bibinfo {author} {\bibfnamefont {C.}~\bibnamefont {Kissling}},
  \bibinfo {author} {\bibfnamefont {V.}~\bibnamefont {Gaydamachenko}}, \bibinfo
  {author} {\bibfnamefont {L.}~\bibnamefont {Gr\"{u}nhaupt}},\ and\ \bibinfo
  {author} {\bibfnamefont {S.}~\bibnamefont {Lotkhov}},\ }\bibfield  {title}
  {\bibinfo {title} {Demonstration of dual {S}hapiro steps in small {J}osephson
  junctions},\ }\href
  {https://doi.org/https://doi.org/10.1038/s41467-024-53011-z} {\bibfield
  {journal} {\bibinfo  {journal} {Nat. Commun.}\ }\textbf {\bibinfo {volume}
  {15}},\ \bibinfo {pages} {8726} (\bibinfo {year}
  {2024}{\natexlab{b}})}\BibitemShut {NoStop}%
\bibitem [{\citenamefont {Crescini}\ \emph {et~al.}(2023)\citenamefont
  {Crescini}, \citenamefont {Cailleaux}, \citenamefont {Guichard},
  \citenamefont {Naud}, \citenamefont {Buisson}, \citenamefont {Murch},\ and\
  \citenamefont {Roch}}]{Crescini2023}%
  \BibitemOpen
  \bibfield  {author} {\bibinfo {author} {\bibfnamefont {N.}~\bibnamefont
  {Crescini}}, \bibinfo {author} {\bibfnamefont {S.}~\bibnamefont {Cailleaux}},
  \bibinfo {author} {\bibfnamefont {W.}~\bibnamefont {Guichard}}, \bibinfo
  {author} {\bibfnamefont {C.}~\bibnamefont {Naud}}, \bibinfo {author}
  {\bibfnamefont {O.}~\bibnamefont {Buisson}}, \bibinfo {author} {\bibfnamefont
  {K.~W.}\ \bibnamefont {Murch}},\ and\ \bibinfo {author} {\bibfnamefont
  {N.}~\bibnamefont {Roch}},\ }\bibfield  {title} {\bibinfo {title} {Evidence
  of dual {S}hapiro steps in a {J}osephson junction array},\ }\href
  {https://doi.org/https://doi.org/10.1038/s41567-023-01961-4} {\bibfield
  {journal} {\bibinfo  {journal} {Nat. Phys.}\ }\textbf {\bibinfo {volume}
  {19}},\ \bibinfo {pages} {851} (\bibinfo {year} {2023})}\BibitemShut
  {NoStop}%
\bibitem [{\citenamefont {Shaikhaidarov}\ \emph
  {et~al.}(2024{\natexlab{a}})\citenamefont {Shaikhaidarov}, \citenamefont
  {Kim}, \citenamefont {Dunstan}, \citenamefont {Antonov}, \citenamefont
  {Antonov},\ and\ \citenamefont {Astafiev}}]{ShaikhaidarovNatCom2024}%
  \BibitemOpen
  \bibfield  {author} {\bibinfo {author} {\bibfnamefont {R.~S.}\ \bibnamefont
  {Shaikhaidarov}}, \bibinfo {author} {\bibfnamefont {K.~H.}\ \bibnamefont
  {Kim}}, \bibinfo {author} {\bibfnamefont {J.}~\bibnamefont {Dunstan}},
  \bibinfo {author} {\bibfnamefont {I.}~\bibnamefont {Antonov}}, \bibinfo
  {author} {\bibfnamefont {V.~N.}\ \bibnamefont {Antonov}},\ and\ \bibinfo
  {author} {\bibfnamefont {O.~V.}\ \bibnamefont {Astafiev}},\ }\bibfield
  {title} {\bibinfo {title} {Quantized current steps due to the synchronization
  of microwaves with {B}loch oscillations in small {J}osephson junctions},\
  }\href {https://doi.org/https://doi.org/10.1038/s41467-024-53600-y}
  {\bibfield  {journal} {\bibinfo  {journal} {Nat. Commun.}\ }\textbf {\bibinfo
  {volume} {15}},\ \bibinfo {pages} {9326} (\bibinfo {year}
  {2024}{\natexlab{a}})}\BibitemShut {NoStop}%
\bibitem [{\citenamefont {Shaikhaidarov}\ \emph {et~al.}(2022)\citenamefont
  {Shaikhaidarov}, \citenamefont {Kim}, \citenamefont {Dunstan}, \citenamefont
  {Antonov}, \citenamefont {Linzen}, \citenamefont {Ziegler}, \citenamefont
  {Golubev}, \citenamefont {Antonov}, \citenamefont {Il’ichev},\ and\
  \citenamefont {Astafiev}}]{Shaikhaidarov2022}%
  \BibitemOpen
  \bibfield  {author} {\bibinfo {author} {\bibfnamefont {R.~S.}\ \bibnamefont
  {Shaikhaidarov}}, \bibinfo {author} {\bibfnamefont {K.~H.}\ \bibnamefont
  {Kim}}, \bibinfo {author} {\bibfnamefont {J.~W.}\ \bibnamefont {Dunstan}},
  \bibinfo {author} {\bibfnamefont {I.~V.}\ \bibnamefont {Antonov}}, \bibinfo
  {author} {\bibfnamefont {S.}~\bibnamefont {Linzen}}, \bibinfo {author}
  {\bibfnamefont {M.}~\bibnamefont {Ziegler}}, \bibinfo {author} {\bibfnamefont
  {D.~S.}\ \bibnamefont {Golubev}}, \bibinfo {author} {\bibfnamefont {V.~N.}\
  \bibnamefont {Antonov}}, \bibinfo {author} {\bibfnamefont {E.~V.}\
  \bibnamefont {Il’ichev}},\ and\ \bibinfo {author} {\bibfnamefont {O.~V.}\
  \bibnamefont {Astafiev}},\ }\bibfield  {title} {\bibinfo {title} {Quantized
  current steps due to the a.c. coherent quantum phase-slip effect},\ }\href
  {https://doi.org/https://doi.org/10.1038/s41586-022-04947-z} {\bibfield
  {journal} {\bibinfo  {journal} {Nature}\ }\textbf {\bibinfo {volume} {608}},\
  \bibinfo {pages} {45} (\bibinfo {year} {2022})}\BibitemShut {NoStop}%
\bibitem [{\citenamefont {Antonov}\ \emph {et~al.}(2026)\citenamefont
  {Antonov}, \citenamefont {Shaikhaidarov}, \citenamefont {Kim}, \citenamefont
  {Golubev}, \citenamefont {Linzen}, \citenamefont {Il’ichev}, \citenamefont
  {Antonov},\ and\ \citenamefont {Astafiev}}]{Antonov2026}%
  \BibitemOpen
  \bibfield  {author} {\bibinfo {author} {\bibfnamefont {I.}~\bibnamefont
  {Antonov}}, \bibinfo {author} {\bibfnamefont {R.~S.}\ \bibnamefont
  {Shaikhaidarov}}, \bibinfo {author} {\bibfnamefont {K.~H.}\ \bibnamefont
  {Kim}}, \bibinfo {author} {\bibfnamefont {D.}~\bibnamefont {Golubev}},
  \bibinfo {author} {\bibfnamefont {S.}~\bibnamefont {Linzen}}, \bibinfo
  {author} {\bibfnamefont {E.~V.}\ \bibnamefont {Il’ichev}}, \bibinfo
  {author} {\bibfnamefont {V.~N.}\ \bibnamefont {Antonov}},\ and\ \bibinfo
  {author} {\bibfnamefont {O.~V.}\ \bibnamefont {Astafiev}},\ }\bibfield
  {title} {\bibinfo {title} {Demonstration of dual {S}hapiro steps in small
  {J}osephson junctions},\ }\href
  {https://doi.org/https://doi.org/10.1038/s41467-025-67735-z} {\bibfield
  {journal} {\bibinfo  {journal} {Nat. Commun.}\ }\textbf {\bibinfo {volume}
  {17}},\ \bibinfo {pages} {1264} (\bibinfo {year} {2026})}\BibitemShut
  {NoStop}%
\bibitem [{\citenamefont {Likharev}(1986{\natexlab{a}})}]{KKL1986preprint}%
  \BibitemOpen
  \bibfield  {author} {\bibinfo {author} {\bibfnamefont {K.~K.}\ \bibnamefont
  {Likharev}},\ }\href@noop {} {\bibinfo {title} {Coexistence of {B}loch and
  {J}osephson effects, and resistance quantization in small-area layered
  superconducting structures}},\ \bibinfo {howpublished} {M. V. Lomonosov
  Moscow State University, Department of Physics} (\bibinfo {year}
  {1986}{\natexlab{a}}),\ \bibinfo {note} {preprint No. 29/1986}\BibitemShut
  {NoStop}%
\bibitem [{\citenamefont {Likharev}\ and\ \citenamefont
  {Zorin}(1987)}]{KKL-AZ-JJAP1987RQ}%
  \BibitemOpen
  \bibfield  {author} {\bibinfo {author} {\bibfnamefont {K.~K.}\ \bibnamefont
  {Likharev}}\ and\ \bibinfo {author} {\bibfnamefont {A.~B.}\ \bibnamefont
  {Zorin}},\ }\bibfield  {title} {\bibinfo {title} {Simultaneous {B}loch and
  {J}osephson oscillations, and resistance quantization in small
  superconducting double junctions},\ }\href
  {https://doi.org/10.7567/JJAPS.26S3.1407} {\bibfield  {journal} {\bibinfo
  {journal} {Jpn. J. Appl. Phys.}\ }\textbf {\bibinfo {volume} {26}},\ \bibinfo
  {pages} {1407} (\bibinfo {year} {1987})}\BibitemShut {NoStop}%
\bibitem [{\citenamefont {Zorin}(1996)}]{AZ-PRL1996}%
  \BibitemOpen
  \bibfield  {author} {\bibinfo {author} {\bibfnamefont {A.~B.}\ \bibnamefont
  {Zorin}},\ }\bibfield  {title} {\bibinfo {title} {Quantum-limited
  electrometer based on single {C}ooper pair tunneling},\ }\href
  {https://doi.org/10.1103/PhysRevLett.76.4408} {\bibfield  {journal} {\bibinfo
   {journal} {Phys. Rev. Lett.}\ }\textbf {\bibinfo {volume} {76}},\ \bibinfo
  {pages} {4408} (\bibinfo {year} {1996})}\BibitemShut {NoStop}%
\bibitem [{\citenamefont {v.~Klitzing}\ \emph {et~al.}(1980)\citenamefont
  {v.~Klitzing}, \citenamefont {Dorda},\ and\ \citenamefont
  {Pepper}}]{KvKlitzingPRL1980}%
  \BibitemOpen
  \bibfield  {author} {\bibinfo {author} {\bibfnamefont {K.}~\bibnamefont
  {v.~Klitzing}}, \bibinfo {author} {\bibfnamefont {G.}~\bibnamefont {Dorda}},\
  and\ \bibinfo {author} {\bibfnamefont {M.}~\bibnamefont {Pepper}},\
  }\bibfield  {title} {\bibinfo {title} {New method for high-accuracy
  determination of the fine-structure constant based on quantized {H}all
  resistance},\ }\href {https://doi.org/10.1103/PhysRevLett.45.494} {\bibfield
  {journal} {\bibinfo  {journal} {Phys. Rev. Lett.}\ }\textbf {\bibinfo
  {volume} {45}},\ \bibinfo {pages} {494} (\bibinfo {year} {1980})}\BibitemShut
  {NoStop}%
\bibitem [{\citenamefont {v.~Klitzing}(1986)}]{KvKlitzingRMP1986}%
  \BibitemOpen
  \bibfield  {author} {\bibinfo {author} {\bibfnamefont {K.}~\bibnamefont
  {v.~Klitzing}},\ }\bibfield  {title} {\bibinfo {title} {The quantized {H}all
  effect},\ }\href {https://doi.org/10.1103/RevModPhys.58.519} {\bibfield
  {journal} {\bibinfo  {journal} {Rev. Mod. Phys.}\ }\textbf {\bibinfo {volume}
  {58}},\ \bibinfo {pages} {519} (\bibinfo {year} {1986})}\BibitemShut
  {NoStop}%
\bibitem [{\citenamefont {Riwar}\ \emph {et~al.}(2016)\citenamefont {Riwar},
  \citenamefont {Houzet}, \citenamefont {Meyer},\ and\ \citenamefont
  {Nazarov}}]{Riwar2016}%
  \BibitemOpen
  \bibfield  {author} {\bibinfo {author} {\bibfnamefont {R.-P.}\ \bibnamefont
  {Riwar}}, \bibinfo {author} {\bibfnamefont {M.}~\bibnamefont {Houzet}},
  \bibinfo {author} {\bibfnamefont {J.~S.}\ \bibnamefont {Meyer}},\ and\
  \bibinfo {author} {\bibfnamefont {Y.~V.}\ \bibnamefont {Nazarov}},\
  }\bibfield  {title} {\bibinfo {title} {Multi-terminal {J}osephson junctions
  as topological matter},\ }\href
  {https://doi.org/https://doi.org/10.1038/ncomms11167} {\bibfield  {journal}
  {\bibinfo  {journal} {Nat. Commun.}\ }\textbf {\bibinfo {volume} {7}},\
  \bibinfo {pages} {11167} (\bibinfo {year} {2016})}\BibitemShut {NoStop}%
\bibitem [{\citenamefont {Eriksson}\ \emph {et~al.}(2017)\citenamefont
  {Eriksson}, \citenamefont {Riwar}, \citenamefont {Houzet}, \citenamefont
  {Meyer},\ and\ \citenamefont {Nazarov}}]{Eriksson2017}%
  \BibitemOpen
  \bibfield  {author} {\bibinfo {author} {\bibfnamefont {E.}~\bibnamefont
  {Eriksson}}, \bibinfo {author} {\bibfnamefont {R.-P.}\ \bibnamefont {Riwar}},
  \bibinfo {author} {\bibfnamefont {M.}~\bibnamefont {Houzet}}, \bibinfo
  {author} {\bibfnamefont {J.~S.}\ \bibnamefont {Meyer}},\ and\ \bibinfo
  {author} {\bibfnamefont {Y.~V.}\ \bibnamefont {Nazarov}},\ }\bibfield
  {title} {\bibinfo {title} {Topological transconductance quantization in a
  four-terminal {J}osephson junction},\ }\href
  {https://doi.org/10.1103/PhysRevB.95.075417} {\bibfield  {journal} {\bibinfo
  {journal} {Phys. Rev. B}\ }\textbf {\bibinfo {volume} {95}},\ \bibinfo
  {pages} {075417} (\bibinfo {year} {2017})}\BibitemShut {NoStop}%
\bibitem [{\citenamefont {Hriscu}\ and\ \citenamefont
  {Nazarov}(2013)}]{HriscuNazarovPRL2013RQ}%
  \BibitemOpen
  \bibfield  {author} {\bibinfo {author} {\bibfnamefont {A.~M.}\ \bibnamefont
  {Hriscu}}\ and\ \bibinfo {author} {\bibfnamefont {V.~V.}\ \bibnamefont
  {Nazarov}},\ }\bibfield  {title} {\bibinfo {title} {Quantum synchronization
  of conjugated variables in a superconducting device leads to the fundamental
  resistance quantization},\ }\href
  {https://doi.org/10.1103/PhysRevLett.110.097002} {\bibfield  {journal}
  {\bibinfo  {journal} {Phys. Rev. Lett.}\ }\textbf {\bibinfo {volume} {110}},\
  \bibinfo {pages} {097002} (\bibinfo {year} {2013})}\BibitemShut {NoStop}%
\bibitem [{\citenamefont {Shaikhaidarov}\ \emph
  {et~al.}(2024{\natexlab{b}})\citenamefont {Shaikhaidarov}, \citenamefont
  {Antonov}, \citenamefont {Kim}, \citenamefont {Shesterikov}, \citenamefont
  {Linzen}, \citenamefont {Il'ichev}, \citenamefont {Antonov},\ and\
  \citenamefont {Astafiev}}]{ShaikhaidarovAPL2024}%
  \BibitemOpen
  \bibfield  {author} {\bibinfo {author} {\bibfnamefont {R.~S.}\ \bibnamefont
  {Shaikhaidarov}}, \bibinfo {author} {\bibfnamefont {I.}~\bibnamefont
  {Antonov}}, \bibinfo {author} {\bibfnamefont {K.~H.}\ \bibnamefont {Kim}},
  \bibinfo {author} {\bibfnamefont {A.}~\bibnamefont {Shesterikov}}, \bibinfo
  {author} {\bibfnamefont {S.}~\bibnamefont {Linzen}}, \bibinfo {author}
  {\bibfnamefont {E.~V.}\ \bibnamefont {Il'ichev}}, \bibinfo {author}
  {\bibfnamefont {V.~N.}\ \bibnamefont {Antonov}},\ and\ \bibinfo {author}
  {\bibfnamefont {O.~V.}\ \bibnamefont {Astafiev}},\ }\bibfield  {title}
  {\bibinfo {title} {Feasibility of the {J}osephson voltage and current
  standards on a single chip},\ }\href {https://doi.org/10.1063/5.0221404}
  {\bibfield  {journal} {\bibinfo  {journal} {Appl. Phys. Lett.}\ }\textbf
  {\bibinfo {volume} {125}},\ \bibinfo {pages} {122602} (\bibinfo {year}
  {2024}{\natexlab{b}})}\BibitemShut {NoStop}%
\bibitem [{\citenamefont {Likharev}(1986{\natexlab{b}})}]{KKLikharev-book}%
  \BibitemOpen
  \bibfield  {author} {\bibinfo {author} {\bibfnamefont {K.~K.}\ \bibnamefont
  {Likharev}},\ }\href@noop {} {\emph {\bibinfo {title} {Dynamics of Josephson
  junctions and circuits}}}\ (\bibinfo  {publisher} {Gordon and Breach},\
  \bibinfo {address} {New York},\ \bibinfo {year} {1986})\BibitemShut {NoStop}%
\bibitem [{\citenamefont {McCumber}(1968)}]{McCumber}%
  \BibitemOpen
  \bibfield  {author} {\bibinfo {author} {\bibfnamefont {D.~E.}\ \bibnamefont
  {McCumber}},\ }\bibfield  {title} {\bibinfo {title} {Effect of ac impedance
  on dc voltage-current characteristics of superconductor weak-link
  junctions},\ }\href {https://doi.org/https://doi.org/10.1063/1.1656743}
  {\bibfield  {journal} {\bibinfo  {journal} {J. Appl. Phys.}\ }\textbf
  {\bibinfo {volume} {39}},\ \bibinfo {pages} {3113} (\bibinfo {year}
  {1968})}\BibitemShut {NoStop}%
\bibitem [{\citenamefont {Stewart}(1968)}]{Stewart}%
  \BibitemOpen
  \bibfield  {author} {\bibinfo {author} {\bibfnamefont {W.~C.}\ \bibnamefont
  {Stewart}},\ }\bibfield  {title} {\bibinfo {title} {Current-voltage
  characteristics of {J}osephson junctions},\ }\href
  {https://doi.org/https://doi.org/10.1063/1.1651991} {\bibfield  {journal}
  {\bibinfo  {journal} {Appl. Phys. Lett.}\ }\textbf {\bibinfo {volume} {12}},\
  \bibinfo {pages} {277} (\bibinfo {year} {1968})}\BibitemShut {NoStop}%
\bibitem [{\citenamefont {Aumentado}\ \emph {et~al.}(2004)\citenamefont
  {Aumentado}, \citenamefont {Keller}, \citenamefont {Martinis},\ and\
  \citenamefont {Devoret}}]{AumentadoPRL2004}%
  \BibitemOpen
  \bibfield  {author} {\bibinfo {author} {\bibfnamefont {J.}~\bibnamefont
  {Aumentado}}, \bibinfo {author} {\bibfnamefont {M.~W.}\ \bibnamefont
  {Keller}}, \bibinfo {author} {\bibfnamefont {J.~M.}\ \bibnamefont
  {Martinis}},\ and\ \bibinfo {author} {\bibfnamefont {M.~H.}\ \bibnamefont
  {Devoret}},\ }\bibfield  {title} {\bibinfo {title} {Nonequilibrium
  quasiparticles and $2e$ periodicity in single-{C}ooper-pair transistors},\
  }\href {https://doi.org/10.1103/PhysRevLett.92.066802} {\bibfield  {journal}
  {\bibinfo  {journal} {Phys. Rev. Lett.}\ }\textbf {\bibinfo {volume} {92}},\
  \bibinfo {pages} {066802} (\bibinfo {year} {2004})}\BibitemShut {NoStop}%
\bibitem [{\citenamefont {Nguyen}\ \emph {et~al.}(2013)\citenamefont {Nguyen},
  \citenamefont {Aref}, \citenamefont {Kauppila}, \citenamefont {Meschke},
  \citenamefont {Winkelmann}, \citenamefont {Courtois},\ and\ \citenamefont
  {Pekola}}]{Nguyen_2013}%
  \BibitemOpen
  \bibfield  {author} {\bibinfo {author} {\bibfnamefont {H.~Q.}\ \bibnamefont
  {Nguyen}}, \bibinfo {author} {\bibfnamefont {T.}~\bibnamefont {Aref}},
  \bibinfo {author} {\bibfnamefont {V.~J.}\ \bibnamefont {Kauppila}}, \bibinfo
  {author} {\bibfnamefont {M.}~\bibnamefont {Meschke}}, \bibinfo {author}
  {\bibfnamefont {C.~B.}\ \bibnamefont {Winkelmann}}, \bibinfo {author}
  {\bibfnamefont {H.}~\bibnamefont {Courtois}},\ and\ \bibinfo {author}
  {\bibfnamefont {J.~P.}\ \bibnamefont {Pekola}},\ }\bibfield  {title}
  {\bibinfo {title} {Trapping hot quasi-particles in a high-power
  superconducting electronic cooler},\ }\href
  {https://doi.org/10.1088/1367-2630/15/8/085013} {\bibfield  {journal}
  {\bibinfo  {journal} {New Journal of Physics}\ }\textbf {\bibinfo {volume}
  {15}},\ \bibinfo {pages} {085013} (\bibinfo {year} {2013})}\BibitemShut
  {NoStop}%
\bibitem [{\citenamefont {Iaia}\ \emph {et~al.}(2022)\citenamefont {Iaia},
  \citenamefont {Ku}, \citenamefont {Ballard}, \citenamefont {Larson},
  \citenamefont {Yelton}, \citenamefont {Liu}, \citenamefont {Patel},
  \citenamefont {McDermott},\ and\ \citenamefont {Plourde}}]{Iaia2022}%
  \BibitemOpen
  \bibfield  {author} {\bibinfo {author} {\bibfnamefont {V.}~\bibnamefont
  {Iaia}}, \bibinfo {author} {\bibfnamefont {J.}~\bibnamefont {Ku}}, \bibinfo
  {author} {\bibfnamefont {A.}~\bibnamefont {Ballard}}, \bibinfo {author}
  {\bibfnamefont {C.~P.}\ \bibnamefont {Larson}}, \bibinfo {author}
  {\bibfnamefont {E.}~\bibnamefont {Yelton}}, \bibinfo {author} {\bibfnamefont
  {C.~H.}\ \bibnamefont {Liu}}, \bibinfo {author} {\bibfnamefont
  {S.}~\bibnamefont {Patel}}, \bibinfo {author} {\bibfnamefont
  {R.}~\bibnamefont {McDermott}},\ and\ \bibinfo {author} {\bibfnamefont
  {B.~L.~T.}\ \bibnamefont {Plourde}},\ }\bibfield  {title} {\bibinfo {title}
  {Phonon downconversion to suppress correlated errors in superconducting
  qubits},\ }\href {https://doi.org/https://doi.org/10.1038/s41467-022-33997-0}
  {\bibfield  {journal} {\bibinfo  {journal} {Nat. Commun.}\ }\textbf {\bibinfo
  {volume} {13}},\ \bibinfo {pages} {6425} (\bibinfo {year}
  {2022})}\BibitemShut {NoStop}%
\bibitem [{\citenamefont {Zorin}(2001)}]{ZorinPRL2001}%
  \BibitemOpen
  \bibfield  {author} {\bibinfo {author} {\bibfnamefont {A.~B.}\ \bibnamefont
  {Zorin}},\ }\bibfield  {title} {\bibinfo {title} {Radio-frequency
  {B}loch-transistor electrometer},\ }\href
  {https://doi.org/10.1103/PhysRevLett.86.3388} {\bibfield  {journal} {\bibinfo
   {journal} {Phys. Rev. Lett.}\ }\textbf {\bibinfo {volume} {86}},\ \bibinfo
  {pages} {3388} (\bibinfo {year} {2001})}\BibitemShut {NoStop}%
\bibitem [{\citenamefont {Vion}\ \emph {et~al.}(2002)\citenamefont {Vion},
  \citenamefont {Aassime}, \citenamefont {Cottet}, \citenamefont {Joyez},
  \citenamefont {Pothier}, \citenamefont {Urbina}, \citenamefont {Esteve},\
  and\ \citenamefont {Devoret}}]{Vion2002}%
  \BibitemOpen
  \bibfield  {author} {\bibinfo {author} {\bibfnamefont {D.}~\bibnamefont
  {Vion}}, \bibinfo {author} {\bibfnamefont {A.}~\bibnamefont {Aassime}},
  \bibinfo {author} {\bibfnamefont {A.}~\bibnamefont {Cottet}}, \bibinfo
  {author} {\bibfnamefont {P.}~\bibnamefont {Joyez}}, \bibinfo {author}
  {\bibfnamefont {H.}~\bibnamefont {Pothier}}, \bibinfo {author} {\bibfnamefont
  {C.}~\bibnamefont {Urbina}}, \bibinfo {author} {\bibfnamefont
  {D.}~\bibnamefont {Esteve}},\ and\ \bibinfo {author} {\bibfnamefont {M.~H.}\
  \bibnamefont {Devoret}},\ }\bibfield  {title} {\bibinfo {title} {Manipulating
  the quantum state of an electrical circuit},\ }\href
  {https://doi.org/10.1126/science.1069372} {\bibfield  {journal} {\bibinfo
  {journal} {Science}\ }\textbf {\bibinfo {volume} {296}},\ \bibinfo {pages}
  {886} (\bibinfo {year} {2002})}\BibitemShut {NoStop}%
\bibitem [{\citenamefont {Zorin}(2002)}]{AZ-PhysC2002}%
  \BibitemOpen
  \bibfield  {author} {\bibinfo {author} {\bibfnamefont {A.~B.}\ \bibnamefont
  {Zorin}},\ }\bibfield  {title} {\bibinfo {title} {Cooper-pair qubit and
  {C}ooper-pair electrometer in one device},\ }\href
  {https://doi.org/https://doi.org/10.1016/S0921-4534(01)01182-0} {\bibfield
  {journal} {\bibinfo  {journal} {Physica C}\ }\textbf {\bibinfo {volume}
  {368}},\ \bibinfo {pages} {284} (\bibinfo {year} {2002})}\BibitemShut
  {NoStop}%
\bibitem [{\citenamefont {Zorin}(2004)}]{AZ-JETP2004}%
  \BibitemOpen
  \bibfield  {author} {\bibinfo {author} {\bibfnamefont {A.~B.}\ \bibnamefont
  {Zorin}},\ }\bibfield  {title} {\bibinfo {title} {Josephson charge-phase
  qubit with the radio frequency readout: coupling and decoherence},\ }\href
  {https://doi.org/https://doi.org/10.1134/1.1777638} {\bibfield  {journal}
  {\bibinfo  {journal} {Zh. Eksp. Teor. Fiz.}\ }\textbf {\bibinfo {volume}
  {125}},\ \bibinfo {pages} {1423} (\bibinfo {year} {2004})},\ \bibinfo {note}
  {[JETP \textbf{98}, 1250-1261 (2004)]}\BibitemShut {NoStop}%
\bibitem [{\citenamefont {Barone}\ and\ \citenamefont
  {Paterno}(1982)}]{Barone-Paterno-book}%
  \BibitemOpen
  \bibfield  {author} {\bibinfo {author} {\bibfnamefont {A.}~\bibnamefont
  {Barone}}\ and\ \bibinfo {author} {\bibfnamefont {G.}~\bibnamefont
  {Paterno}},\ }\href@noop {} {\emph {\bibinfo {title} {Physics and
  Applications of the Josephson Effect}}}\ (\bibinfo  {publisher} {John Wiley
  \& Sons Inc.},\ \bibinfo {address} {New York},\ \bibinfo {year}
  {1982})\BibitemShut {NoStop}%
\bibitem [{\citenamefont {Zorin}\ \emph {et~al.}(1998)\citenamefont {Zorin},
  \citenamefont {Pashkin}, \citenamefont {Krupenin},\ and\ \citenamefont
  {Scherer}}]{AZ_Pashkin1998}%
  \BibitemOpen
  \bibfield  {author} {\bibinfo {author} {\bibfnamefont {A.~B.}\ \bibnamefont
  {Zorin}}, \bibinfo {author} {\bibfnamefont {Y.~A.}\ \bibnamefont {Pashkin}},
  \bibinfo {author} {\bibfnamefont {V.~A.}\ \bibnamefont {Krupenin}},\ and\
  \bibinfo {author} {\bibfnamefont {H.}~\bibnamefont {Scherer}},\ }\bibfield
  {title} {\bibinfo {title} {Coulomb blockade electrometer based on single
  {C}ooper pair tunneling},\ }\href
  {https://doi.org/https://doi.org/10.1016/S0964-1807(98)00116-1} {\bibfield
  {journal} {\bibinfo  {journal} {Appl. Supercond.}\ }\textbf {\bibinfo
  {volume} {6}},\ \bibinfo {pages} {453} (\bibinfo {year} {1998})}\BibitemShut
  {NoStop}%
\bibitem [{\citenamefont {Zorin}\ \emph {et~al.}(1999)\citenamefont {Zorin},
  \citenamefont {Lotkhov}, \citenamefont {Pashkin}, \citenamefont {Zangerle},
  \citenamefont {Kruperin}, \citenamefont {Weimann}, \citenamefont {Scherer},\
  and\ \citenamefont {Niemeyer}}]{AZ-1999-BTelectrometers}%
  \BibitemOpen
  \bibfield  {author} {\bibinfo {author} {\bibfnamefont {A.~B.}\ \bibnamefont
  {Zorin}}, \bibinfo {author} {\bibfnamefont {S.~V.}\ \bibnamefont {Lotkhov}},
  \bibinfo {author} {\bibfnamefont {Y.~A.}\ \bibnamefont {Pashkin}}, \bibinfo
  {author} {\bibfnamefont {H.}~\bibnamefont {Zangerle}}, \bibinfo {author}
  {\bibfnamefont {V.~A.}\ \bibnamefont {Kruperin}}, \bibinfo {author}
  {\bibfnamefont {T.}~\bibnamefont {Weimann}}, \bibinfo {author} {\bibfnamefont
  {H.}~\bibnamefont {Scherer}},\ and\ \bibinfo {author} {\bibfnamefont
  {J.}~\bibnamefont {Niemeyer}},\ }\bibfield  {title} {\bibinfo {title} {Highly
  sensitive electrometers based on single {C}ooper pair tunneling},\ }\href
  {https://doi.org/https://doi.org/10.1023/A:1007780925567} {\bibfield
  {journal} {\bibinfo  {journal} {J. Supercond.}\ }\textbf {\bibinfo {volume}
  {12}},\ \bibinfo {pages} {747} (\bibinfo {year} {1999})}\BibitemShut
  {NoStop}%
\bibitem [{\citenamefont {Houzet}\ \emph {et~al.}(2026)\citenamefont {Houzet},
  \citenamefont {Vakhtel},\ and\ \citenamefont {Meyer}}]{houzet2026blochdiode}%
  \BibitemOpen
  \bibfield  {author} {\bibinfo {author} {\bibfnamefont {M.}~\bibnamefont
  {Houzet}}, \bibinfo {author} {\bibfnamefont {T.}~\bibnamefont {Vakhtel}},\
  and\ \bibinfo {author} {\bibfnamefont {J.~S.}\ \bibnamefont {Meyer}},\
  }\bibfield  {title} {\bibinfo {title} {Bloch diode},\ }\href
  {https://doi.org/10.1103/q98p-j4vh} {\bibfield  {journal} {\bibinfo
  {journal} {Phys. Rev. Lett.}\ }\textbf {\bibinfo {volume} {136}},\ \bibinfo
  {pages} {146001} (\bibinfo {year} {2026})}\BibitemShut {NoStop}%
\bibitem [{\citenamefont {Bouchiat}\ \emph {et~al.}(1998)\citenamefont
  {Bouchiat}, \citenamefont {Vion}, \citenamefont {Joyez}, \citenamefont
  {Esteve},\ and\ \citenamefont {Devoret}}]{Bouchiat1998}%
  \BibitemOpen
  \bibfield  {author} {\bibinfo {author} {\bibfnamefont {V.}~\bibnamefont
  {Bouchiat}}, \bibinfo {author} {\bibfnamefont {D.}~\bibnamefont {Vion}},
  \bibinfo {author} {\bibfnamefont {P.}~\bibnamefont {Joyez}}, \bibinfo
  {author} {\bibfnamefont {D.}~\bibnamefont {Esteve}},\ and\ \bibinfo {author}
  {\bibfnamefont {M.~H.}\ \bibnamefont {Devoret}},\ }\bibfield  {title}
  {\bibinfo {title} {Quantum coherence with a single {C}ooper pair},\ }\href
  {https://doi.org/10.1238/Physica.Topical.076a00165} {\bibfield  {journal}
  {\bibinfo  {journal} {Phys. Scripta}\ }\textbf {\bibinfo {volume} {T76}},\
  \bibinfo {pages} {165} (\bibinfo {year} {1998})}\BibitemShut {NoStop}%
\bibitem [{\citenamefont {Lifshitz}\ and\ \citenamefont
  {Pitaevskii}(1980)}]{LifshitzPitaevskii-book}%
  \BibitemOpen
  \bibfield  {author} {\bibinfo {author} {\bibfnamefont {E.~M.}\ \bibnamefont
  {Lifshitz}}\ and\ \bibinfo {author} {\bibfnamefont {L.~P.}\ \bibnamefont
  {Pitaevskii}},\ }\href@noop {} {\emph {\bibinfo {title} {Statistical Physics,
  Part 2: Theory of the Condensed State (Course of Theoretical Physics Vol.
  9)}}}\ (\bibinfo  {publisher} {Pergamon Press / Butterworth-Heinemann},
  \bibinfo {address} {Oxford},\
  \bibinfo {year} {1980})\BibitemShut {NoStop}%
\bibitem [{\citenamefont {Matveev}\ \emph {et~al.}(1993)\citenamefont
  {Matveev}, \citenamefont {Gisself\"alt}, \citenamefont {Glazman},
  \citenamefont {Jonson},\ and\ \citenamefont {Shekhter}}]{Matveev1993}%
  \BibitemOpen
  \bibfield  {author} {\bibinfo {author} {\bibfnamefont {K.~A.}\ \bibnamefont
  {Matveev}}, \bibinfo {author} {\bibfnamefont {M.}~\bibnamefont
  {Gisself\"alt}}, \bibinfo {author} {\bibfnamefont {L.~I.}\ \bibnamefont
  {Glazman}}, \bibinfo {author} {\bibfnamefont {M.}~\bibnamefont {Jonson}},\
  and\ \bibinfo {author} {\bibfnamefont {R.~I.}\ \bibnamefont {Shekhter}},\
  }\bibfield  {title} {\bibinfo {title} {Parity-induced suppression of the
  {C}oulomb blockade of {J}osephson tunneling},\ }\href
  {https://doi.org/10.1103/PhysRevLett.70.2940} {\bibfield  {journal} {\bibinfo
   {journal} {Phys. Rev. Lett.}\ }\textbf {\bibinfo {volume} {70}},\ \bibinfo
  {pages} {2940} (\bibinfo {year} {1993})}\BibitemShut {NoStop}%
\bibitem [{\citenamefont {Joyez}\ \emph {et~al.}(1994)\citenamefont {Joyez},
  \citenamefont {Lafarge}, \citenamefont {Filipe}, \citenamefont {Esteve},\
  and\ \citenamefont {Devoret}}]{Joyez1994}%
  \BibitemOpen
  \bibfield  {author} {\bibinfo {author} {\bibfnamefont {P.}~\bibnamefont
  {Joyez}}, \bibinfo {author} {\bibfnamefont {P.}~\bibnamefont {Lafarge}},
  \bibinfo {author} {\bibfnamefont {A.}~\bibnamefont {Filipe}}, \bibinfo
  {author} {\bibfnamefont {D.}~\bibnamefont {Esteve}},\ and\ \bibinfo {author}
  {\bibfnamefont {M.~H.}\ \bibnamefont {Devoret}},\ }\bibfield  {title}
  {\bibinfo {title} {Observation of parity-induced suppression of {J}osephson
  tunneling in the superconducting single electron transistor},\ }\href
  {https://doi.org/10.1103/PhysRevLett.72.2458} {\bibfield  {journal} {\bibinfo
   {journal} {Phys. Rev. Lett.}\ }\textbf {\bibinfo {volume} {72}},\ \bibinfo
  {pages} {2458} (\bibinfo {year} {1994})}\BibitemShut {NoStop}%
\bibitem [{\citenamefont {Eiles}\ and\ \citenamefont
  {Martinis}(1994)}]{EilesMartinis1994}%
  \BibitemOpen
  \bibfield  {author} {\bibinfo {author} {\bibfnamefont {T.~M.}\ \bibnamefont
  {Eiles}}\ and\ \bibinfo {author} {\bibfnamefont {J.~M.}\ \bibnamefont
  {Martinis}},\ }\bibfield  {title} {\bibinfo {title} {Combined {J}osephson and
  charging behavior of the supercurrent in the superconducting single-electron
  transistor},\ }\href {https://doi.org/10.1103/PhysRevB.50.627} {\bibfield
  {journal} {\bibinfo  {journal} {Phys. Rev. B}\ }\textbf {\bibinfo {volume}
  {50}},\ \bibinfo {pages} {627} (\bibinfo {year} {1994})}\BibitemShut
  {NoStop}%
\bibitem [{\citenamefont {Migulin}\ \emph {et~al.}(1983)\citenamefont
  {Migulin}, \citenamefont {Medvedev}, \citenamefont {Mustel},\ and\
  \citenamefont {Parygin}}]{Migulin-book}%
  \BibitemOpen
  \bibfield  {author} {\bibinfo {author} {\bibfnamefont {V.~V.}\ \bibnamefont
  {Migulin}}, \bibinfo {author} {\bibfnamefont {V.~I.}\ \bibnamefont
  {Medvedev}}, \bibinfo {author} {\bibfnamefont {E.~R.}\ \bibnamefont
  {Mustel}},\ and\ \bibinfo {author} {\bibfnamefont {V.~N.}\ \bibnamefont
  {Parygin}},\ }\href@noop {} {\emph {\bibinfo {title} {Basic Theory of
  Oscillations}}}\ (\bibinfo  {publisher} {Mir},\ \bibinfo {address} {Moscow},\
  \bibinfo {year} {1983})\BibitemShut {NoStop}%
\bibitem [{\citenamefont {Akhmanov}\ \emph {et~al.}(1981)\citenamefont
  {Akhmanov}, \citenamefont {Djakov},\ and\ \citenamefont
  {Chirkin}}]{Axmanov-book}%
  \BibitemOpen
  \bibfield  {author} {\bibinfo {author} {\bibfnamefont {S.~A.}\ \bibnamefont
  {Akhmanov}}, \bibinfo {author} {\bibfnamefont {Y.~E.}\ \bibnamefont
  {Djakov}},\ and\ \bibinfo {author} {\bibfnamefont {A.~S.}\ \bibnamefont
  {Chirkin}},\ }\href@noop {} {\emph {\bibinfo {title} {Vvedenije v
  statisticheskuju radiofiziku i optiku (Introduction to Statistical
  Radiophysics and Optics)}}}\ (\bibinfo  {publisher} {Nauka},\ \bibinfo
  {address} {Moscow},\ \bibinfo {year} {1981})\ \bibinfo {note} {[In
  Russian]}\BibitemShut {NoStop}%
\bibitem [{\citenamefont {Zorin}(2026)}]{Supp_Info_Zenodo2026}%
  \BibitemOpen
  \bibfield  {author} {\bibinfo {author} {\bibfnamefont {A.~B.}\ \bibnamefont
  {Zorin}},\ } {Additional information for the work {\enquote{{D}ual {S}hapiro steps and fundamental 
  transconductance in the dc-driven {B}loch transistor.}}  
  \href {https://doi.org/10.5281/zenodo.20277445} {\bibinfo {title}
  {Z}enodo. https://doi.org/10.5281/zenodo.20277445}} (\bibinfo
  {year} {2026})\BibitemShut {NoStop}%
\bibitem [{\citenamefont {Vanneste}\ \emph {et~al.}(1981)\citenamefont
  {Vanneste}, \citenamefont {Gilabert}, \citenamefont {Sibillot},\ and\
  \citenamefont {Ostrowsky}}]{Vanneste-JLTP81}%
  \BibitemOpen
  \bibfield  {author} {\bibinfo {author} {\bibfnamefont {C.}~\bibnamefont
  {Vanneste}}, \bibinfo {author} {\bibfnamefont {A.}~\bibnamefont {Gilabert}},
  \bibinfo {author} {\bibfnamefont {P.}~\bibnamefont {Sibillot}},\ and\
  \bibinfo {author} {\bibfnamefont {D.~B.}\ \bibnamefont {Ostrowsky}},\
  }\bibfield  {title} {\bibinfo {title} {Josephson steps produced by critical
  current amplitude modulation},\ }\href
  {https://doi.org/https://doi.org/10.1007/BF00654498} {\bibfield  {journal}
  {\bibinfo  {journal} {J. Low Temp. Phys.}\ }\textbf {\bibinfo {volume}
  {45}},\ \bibinfo {pages} {517-530} (\bibinfo {year} {1981})}\BibitemShut
  {NoStop}%
\bibitem [{\citenamefont {Vanneste}\ \emph {et~al.}(1988)\citenamefont
  {Vanneste}, \citenamefont {Chi}, \citenamefont {Gallagher}, \citenamefont
  {Kleinsasser}, \citenamefont {Raider},\ and\ \citenamefont
  {Sandstrom}}]{Vanneste-JAP88}%
  \BibitemOpen
  \bibfield  {author} {\bibinfo {author} {\bibfnamefont {C.}~\bibnamefont
  {Vanneste}}, \bibinfo {author} {\bibfnamefont {C.~C.}\ \bibnamefont {Chi}},
  \bibinfo {author} {\bibfnamefont {W.~J.}\ \bibnamefont {Gallagher}}, \bibinfo
  {author} {\bibfnamefont {A.~W.}\ \bibnamefont {Kleinsasser}}, \bibinfo
  {author} {\bibfnamefont {S.~I.}\ \bibnamefont {Raider}},\ and\ \bibinfo
  {author} {\bibfnamefont {R.~L.}\ \bibnamefont {Sandstrom}},\ }\bibfield
  {title} {\bibinfo {title} {Shapiro steps on currentvoltage curves of dc
  {SQUID}s},\ }\href {https://doi.org/https://doi.org/10.1063/1.341471}
  {\bibfield  {journal} {\bibinfo  {journal} {J. Appl. Phys.}\ }\textbf
  {\bibinfo {volume} {64}},\ \bibinfo {pages} {242} (\bibinfo {year}
  {1988})}\BibitemShut {NoStop}%
\bibitem [{\citenamefont {Bogoliubov}\ and\ \citenamefont
  {Mitropolski}(1961)}]{Bogoliubov-Mitropolski-book}%
  \BibitemOpen
  \bibfield  {author} {\bibinfo {author} {\bibfnamefont {N.~N.}\ \bibnamefont
  {Bogoliubov}}\ and\ \bibinfo {author} {\bibfnamefont {Y.~A.}\ \bibnamefont
  {Mitropolski}},\ }\href@noop {} {\emph {\bibinfo {title} {Asymptotic Methods
  in the Theory of Non-Linear Oscillations}}}\ (\bibinfo  {publisher} {Gordon
  and Breach},\ \bibinfo {address} {New York},\ \bibinfo {year}
  {1961})\BibitemShut {NoStop}%
\bibitem [{\citenamefont {Stephen}(1969)}]{StephenPhysRev1969}%
  \BibitemOpen
  \bibfield  {author} {\bibinfo {author} {\bibfnamefont {M.~J.}\ \bibnamefont
  {Stephen}},\ }\bibfield  {title} {\bibinfo {title} {Noise in a driven
  {J}osephson oscillator},\ }\href {https://doi.org/10.1103/PhysRev.186.393}
  {\bibfield  {journal} {\bibinfo
  {journal} {Phys. Rev.}\ }\textbf {\bibinfo {volume} {186}},\ \bibinfo {pages}
  {393} (\bibinfo {year} {1969})}\BibitemShut {NoStop}%
\bibitem [{\citenamefont {Marco}\ \emph {et~al.}(2015)\citenamefont {Marco},
  \citenamefont {Hekking},\ and\ \citenamefont {Rastelli}}]{DiMarco2015}%
  \BibitemOpen
  \bibfield  {author} {\bibinfo {author} {\bibfnamefont {A.~D.}\ \bibnamefont
  {Marco}}, \bibinfo {author} {\bibfnamefont {F.~W.~J.}\ \bibnamefont
  {Hekking}},\ and\ \bibinfo {author} {\bibfnamefont {G.}~\bibnamefont
  {Rastelli}},\ }\bibfield  {title} {\bibinfo {title} {Quantum phase-slip
  junction under microwave irradiation},\ }\href
  {https://doi.org/10.1103/PhysRevB.91.184512} {\bibfield  {journal} {\bibinfo
  {journal} {Phys. Rev. B}\ }\textbf {\bibinfo {volume} {91}},\ \bibinfo
  {pages} {184512} (\bibinfo {year} {2015})}\BibitemShut {NoStop}%
\bibitem [{\citenamefont {Kurilovich}\ \emph {et~al.}(2025)\citenamefont
  {Kurilovich}, \citenamefont {Remez},\ and\ \citenamefont
  {Glazman}}]{Kurilovich2025}%
  \BibitemOpen
  \bibfield  {author} {\bibinfo {author} {\bibfnamefont {V.~D.}\ \bibnamefont
  {Kurilovich}}, \bibinfo {author} {\bibfnamefont {B.}~\bibnamefont {Remez}},\
  and\ \bibinfo {author} {\bibfnamefont {L.~I.}\ \bibnamefont {Glazman}},\
  }\bibfield  {title} {\bibinfo {title} {Quantum theory of {B}loch oscillations
  in a resistively shunted transmon},\ }\href
  {https://doi.org/https://doi.org/10.1038/s41467-025-56411-x} {\bibfield
  {journal} {\bibinfo  {journal} {Nat. Commun.}\ }\textbf {\bibinfo {volume}
  {16}},\ \bibinfo {pages} {1384} (\bibinfo {year} {2025})}\BibitemShut
  {NoStop}%
\bibitem [{\citenamefont {Resch}\ \emph {et~al.}(2026)\citenamefont {Resch},
  \citenamefont {Ankerhold}, \citenamefont {Donvil}, \citenamefont
  {Muratore-Ginanneschi},\ and\ \citenamefont
  {Golubev}}]{resch2025shapirosteps}%
  \BibitemOpen
  \bibfield  {author} {\bibinfo {author} {\bibfnamefont {M.}~\bibnamefont
  {Resch}}, \bibinfo {author} {\bibfnamefont {J.}~\bibnamefont {Ankerhold}},
  \bibinfo {author} {\bibfnamefont {B.~I.~C.}\ \bibnamefont {Donvil}}, \bibinfo
  {author} {\bibfnamefont {P.}~\bibnamefont {Muratore-Ginanneschi}},\ and\
  \bibinfo {author} {\bibfnamefont {D.}~\bibnamefont {Golubev}},\ }\bibfield
  {title} {\bibinfo {title} {Conventional and dual {S}hapiro steps in small
  {J}osephson junctions},\ }\href {https://doi.org/10.1103/smby-m2bb}
  {\bibfield  {journal} {\bibinfo  {journal} {Phys. Rev. B}\ }\textbf {\bibinfo
  {volume} {114}},\ \bibinfo {pages} {084501} (\bibinfo {year}
  {2026})}\BibitemShut {NoStop}%
\bibitem [{\citenamefont {Arndt}\ \emph {et~al.}(2018)\citenamefont {Arndt},
  \citenamefont {Roy},\ and\ \citenamefont {Hassler}}]{Arndt2018}%
  \BibitemOpen
  \bibfield  {author} {\bibinfo {author} {\bibfnamefont {L.}~\bibnamefont
  {Arndt}}, \bibinfo {author} {\bibfnamefont {A.}~\bibnamefont {Roy}},\ and\
  \bibinfo {author} {\bibfnamefont {F.}~\bibnamefont {Hassler}},\ }\bibfield
  {title} {\bibinfo {title} {Dual {S}hapiro steps of a phase-slip junction in
  the presence of a parasitic capacitance},\ }\href
  {https://doi.org/10.1103/PhysRevB.98.014525} {\bibfield  {journal} {\bibinfo
  {journal} {Phys. Rev. B}\ }\textbf {\bibinfo {volume} {98}},\ \bibinfo
  {pages} {014525} (\bibinfo {year} {2018})}\BibitemShut {NoStop}%
\bibitem [{\citenamefont {Maibaum}\ \emph {et~al.}(2011)\citenamefont
  {Maibaum}, \citenamefont {Lotkhov},\ and\ \citenamefont
  {Zorin}}]{MaibaumPRB2011}%
  \BibitemOpen
  \bibfield  {author} {\bibinfo {author} {\bibfnamefont {F.}~\bibnamefont
  {Maibaum}}, \bibinfo {author} {\bibfnamefont {S.~V.}\ \bibnamefont
  {Lotkhov}},\ and\ \bibinfo {author} {\bibfnamefont {A.~B.}\ \bibnamefont
  {Zorin}},\ }\bibfield  {title} {\bibinfo {title} {Towards the observation of
  phase-locked {B}loch oscillations in arrays of small {J}osephson junctions},\
  }\href {https://doi.org/10.1103/PhysRevB.84.174514} {\bibfield  {journal}
  {\bibinfo  {journal} {Phys. Rev. B}\ }\textbf {\bibinfo {volume} {84}},\
  \bibinfo {pages} {174514} (\bibinfo {year} {2011})}\BibitemShut {NoStop}%
\bibitem [{\citenamefont {Clarke}\ and\ \citenamefont
  {Braginski}(2004)}]{Clarke-Braginski-book}%
  \BibitemOpen
  \bibfield  {author} {\bibinfo {author} {\bibfnamefont {J.}~\bibnamefont
  {Clarke}}\ and\ \bibinfo {author} {\bibfnamefont {A.}~\bibnamefont
  {Braginski}},\ }\href@noop {} {\emph {\bibinfo {title} {The {SQUID} Handbook:
  Fundamentals and Technology of {SQUID}s and {SQUID} Systems}}}\ (\bibinfo
  {publisher} {Wiley-VCH Verlag GmbH},\ \bibinfo {address} {Berlin},\ \bibinfo
  {year} {2004})\BibitemShut {NoStop}%
\bibitem [{\citenamefont {Likharev}\ and\ \citenamefont
  {Semenov}(1971)}]{LikharevSemenov1971}%
  \BibitemOpen
  \bibfield  {author} {\bibinfo {author} {\bibfnamefont {K.~K.}\ \bibnamefont
  {Likharev}}\ and\ \bibinfo {author} {\bibfnamefont {V.~K.}\ \bibnamefont
  {Semenov}},\ }\bibfield  {title} {\bibinfo {title} {Electrodynamic properties
  of superconducting point contacts},\ }\href@noop {} {\bibfield  {journal}
  {\bibinfo  {journal} {Radiotekhnika i elektronika [in Russian]}\ }\textbf
  {\bibinfo {volume} {16}},\ \bibinfo {pages} {2167} (\bibinfo {year}
  {1971})},\ \bibinfo {note} {[Radio Eng. Electron. Phys. (USSR) (Engl.
  Transl.) \textbf{16}, 1917 (1971)]}\BibitemShut {NoStop}%
\bibitem [{\citenamefont {Wiersma}\ and\ \citenamefont
  {Mana}(2021)}]{Wiersma2021}%
  \BibitemOpen
  \bibfield  {author} {\bibinfo {author} {\bibfnamefont {D.~S.}\ \bibnamefont
  {Wiersma}}\ and\ \bibinfo {author} {\bibfnamefont {G.}~\bibnamefont {Mana}},\
  }\bibfield  {title} {\bibinfo {title} {The fundamental constants of physics
  and the international system of units},\ }\href
  {https://doi.org/10.1007/s12210-021-01022-z} {\bibfield  {journal} {\bibinfo
  {journal} {Rend. Fis. Acc. Lincei}\ }\textbf {\bibinfo {volume} {32}},\
  \bibinfo {pages} {655} (\bibinfo {year} {2021})}\BibitemShut {NoStop}%
\end{thebibliography}
\end{document}